\documentstyle[epsfig]{mn}

\newcommand{\be}{\begin{equation}}
\newcommand{\ee}{\end{equation}}

 \newcommand{\apj}{{\it ApJ, }}
 \newcommand{\apjl}{{\it ApJ(Letters), }}
 
 \newcommand{\aj}{{\it AJ, }}
 \newcommand{\mnr}{{\it MNRAS, }}
 \newcommand{\nat}{{\it Nat, }}

\newcounter{pp3}
\addtocounter{pp3}{3}

\title[The Bardeen-Petterson Effect]{Hydrodynamic Simulations of the  
Bardeen--Petterson Effect}

\author[Richard P. Nelson and John C.B. Papaloizou]{Richard P. Nelson 
\thanks{Email address:
R.P.Nelson@qmw.ac.uk} \& John C.B. Papaloizou \thanks {Email address:
J.C.B.Papaloizou@qmw.ac.uk} \\
 Astronomy Unit, 
 Queen Mary \& Westfield College, Mile End
 Rd, London E1 4NS}

 \date{Received *****; in original form ******}

 \def\LaTeX{L\kern-.36em\raise.3ex\hbox{a}\kern-.15em
	 T\kern-.1667em\lower.7ex\hbox{E}\kern-.125emX}

\begin{document}
 \label{firstpage}

 \maketitle

 \begin{abstract}
We present SPH simulations of accretion discs in orbit about rotating compact
objects, such as black 
holes and neutron stars,
and study the structure of warped discs produced by the 
Bardeen-Petterson
effect.  We calculate the transition radius
out to which the disc specific angular momentum
vector is aligned with that of the black hole.
We focus on the parameter regime 
where the  warp
dynamics are controlled by bending wave propagation, but also consider models
in which warps are subject to diffusion rather than wave transport,
and are
able to consider the fully nonlinear regime.

Because of hydrodynamic or pressure effects,
for the parameter range investigated,
the transition radius is always found to be
much smaller than that obtained by Bardeen \& Petterson (1975).
For discs with midplane Mach  numbers of $\sim 10$, the transition occurs 
between 10 -- 16 R$_+$ (gravitational radii), whereas for a Mach number of 
$\sim 30$ it occurs at around 30 R$_+$. A thicker disc with a Mach number of 5
is found to produce no discernible warped structure.

The rate of black hole -- disc alignment is found to be consistent
with the ideas of Rees (1978), 
with the alignment torque  behaving as  if it arises 
from the accreted material transferring its misaligned 
component of angular momentum at the larger transition 
radius of Bardeen \& Petterson (1975). 

The inclusion of Einstein precession
in the calculations modified both the warped disc
structure and, consistent with linear analysis,
 produced an increased
 alignment rate by up to a factor of 4 because of the
effect that  the non Keplerian 
potential has on the propagation of warps.

 \end{abstract}


\section{Introduction}

Accretion discs  may occur around stellar mass  black holes  in X--ray sources
such as Cygnus X--1, or around  very massive black holes in active galaxies.
The Lense-Thirring effect, or dragging of inertial frames, has
an important influence on the structure of an accretion disc
around a rotating black hole in the situation where
the equatorial plane of the outer disc far from the black
hole is misaligned with the symmetry plane
of the black hole.  

Bardeen \& Petterson (1975)
showed that a viscous disc would be expected to
relax to a form in which the inner regions become aligned
with the equatorial plane of the black hole out to a transition
radius, beyond which the disc remains aligned with the outer 
disc. This phenomenon may be important for understanding
absorption and reprocessing of X--rays in X--ray sources,
and also for providing stability for jet directions in
radio sources (Rees 1978).

In their calculation Bardeen \& Petterson (1975) assumed that
the warps in the misaligned disc diffused
on a viscous time scale. The viscosity coefficient was
taken to be  of the the same  magnitude
as that producing the mass inflow, typically modelled
using the standard Shakura \& Sunyaev (1973) $\alpha$
prescription. However, Papaloizou and Pringle (1983)
showed that  for an isotropic
viscosity,  warp  diffusion occurs at a rate faster by $\sim 1/(2\alpha^2),$
provided $\alpha > H/r,$ much reducing the transition radius.
For  $\alpha < H/r$  it is expected that the dynamics is controlled
by the wave modes present in the disc.

The location of the transition radius has been investigated 
in the  diffusive linear regime ($\alpha > H/r$)
(Kumar \&
Pringle 1985),  and in a model including viscous effects but neglecting
pressure
(Scheuer \& Feiler 1996). The Bardeen--Petterson effect has also been studied 
in the linear regime, but with $\alpha < H/r$, by Ivanov \& Illarianov (1997),
where they also considered the effects of some additional post Newtonian 
corrections.

In this paper we study the Bardeen-Petterson effect
using SPH simulations. We focus primarily
on the parameter regime where the warp
dynamics are controlled by bending wave propagation, although we
also examine models in which warps are subject to diffusion
and are not transmitted by waves. 
The properties
of  the wave and diffusive communication  that occurs
in the discs  modelled here are discussed in detail in a recent paper
(Nelson \& Papaloizou 1999)  hereafter referred to as paper I.
In that paper we compare SPH simulations of small
amplitude disturbances to results obtained using the linearized fluid
equations that are solved using 
a standard finite difference code, though we comment more generally that
by using SPH calculations we are able to consider the fully non linear
regime.
The calculations presented in paper I are used to
calibrate the code viscosity. The warping of a gaseous disc generates
vertical shearing motions ({\em i.e.} horizontal motions that are a function
of the vertical position in the disc). The viscosity that acts on this
vertical shear may not be the same as that which acts on the usual horizontal
shear associated with a planar, differentially rotating accretion 
disc. We therefore
denote the dimensionless viscosity parameter acting on the vertical shear
as $\alpha_1$, and that acting on the horizontal shear through the
$r$--$\phi$ component of the stress tensor as $\alpha$. 
The test calculations presented in paper I indicate that the thicker discs
that we consider, with midplane Mach numbers of ${\cal M} \sim 12$, 
(where $H/r \sim {\cal M}^{-1}$) have
$\alpha_1 \simeq 0.04$ so that bending waves are expected to propagate in these
discs. We note that calibration runs indicate that
$\alpha \simeq 0.02$ -- $0.03$ for disc models of this thickness obtained 
using SPH (e.g. Larwood {\em et al.} 1996; Bryden {\em et al.} 1999).
It is found that the thinner discs with ${\cal M} \sim 30$ have
values of $\alpha_1 \sim 0.1$ and similar values for $\alpha$
so that in these cases warping disturbances
are expected to evolve diffusively.

We  study the
Bardeen--Petterson effect for a variety of disc Mach numbers,
and find that the transition radius is always 
much smaller than that obtained by Bardeen \& Petterson (1975). 
The rate of black hole -- disc alignment is, however, 
found to be consistent
with the  ideas of Rees (1978), and the later work of  Scheuer \& Feiler (1996)
when taking the viscous diffusion coefficients acting in and out of the
plane to be equal. 
A discussion of this result is presented in section~\ref{align}.

The plan of the paper is as follows. In section 2 we give the basic equations.
In section 3 we describe how Lense-Thirring precession
acting on a misaligned accretion disc
in the neighbourhood of a rotating black hole leads
to the Bardeen-Petterson effect, 
where the inner regions  of the disc
become aligned with the  equatorial plane of the black hole 
out to  the transition radius.

The  relaxation of an accretion  disc for which the
angular momentum vector is misaligned with the spin vector of a
rotating black hole
is dependent on the dynamics of warps or bending 
waves (see paper I and references therein).
Linear bending waves are expected to propagate if the 
Shakura \& Sunyaev (1973) $\alpha$ viscosity
appropriate to dissipation of vertical shear, $\alpha_1 < H/r.$
For larger values of  $\alpha_1,$ the evolution is diffusive.
In section 4 we consider the relaxation of a disc subject
to gravomagnetic forces when there is a small misalignment.

In section 5 we discuss  estimates of the Bardeen-Petterson
transition radius  based on equating the Lense-Thirring precessional shear rate
to the viscous diffusion rate. 
Here we note that it is the diffusion rate of warps that is important
and this occurs at a rate faster by $\sim 1/(2\alpha^2)$ than the standard
viscous rate (Papaloizou \& Pringle 1983)  if an isotropic standard Shakura \&
Sunyaev (1973) 
$\alpha$ viscosity is used so $\alpha =\alpha_1$.  The transition radius then
decreases as $\alpha$ decreases  until  $\alpha = H/r.$ Then the warp diffusion rate
becomes the same as the  propagation rate of bending waves
and the transition radius cannot decrease further
as $\alpha$ is reduced.  

In section 6 we describe the implementation of SPH used to
perform the simulations,
and in section 7 we  describe the calculations of the Bardeen-Petterson effect. 
In section 8 we  present the results of
nonlinear simulations of a disc orbiting around a
rotating black hole with Kerr parameter $a=1$
and which has its spin vector initially misaligned
with that of the disc. We use the lowest order post
Newtonian approximation in our simulations.  

We present simulations for Mach numbers ranging between 5 and 30
and inclinations of the outer disc plane of 10 and 30 degrees.
For the effective viscosity  $\alpha_1$ in the disc,
in the lower Mach number cases, warps 
are governed by bending waves,  while for the higher Mach number
case of 30 they are diffusive.
 
The central portion
of the disc becomes aligned with the equatorial plane of the hole
out to a transition radius
which was found to be much smaller than that given by 
Bardeen \& Petterson (1975),
ranging between 15 and 30 gravitational radii.

Models were considered both at high and low inclination. In the nonlinear
high inclination case
the transition between the aligned and non-aligned disc  was more abrupt
than in the low inclination case, indicative of
a tendency for the outer part of the disc to become disconnected
from the inner part.

A calculation is presented for a disc of substantially larger outer
radius than the other disc models we consider. We do not expect that the
location of the transition radius will be affected by the location of the outer
disc radius provided that it is sufficiently far away from the transition zone.
In the majority of cases that we consider, the outer disc radius ($r \sim 2$)
is 
a factor of between three and four times larger than the final
transition radius ($r \sim 0.7$).
We find that a disc that is more than a factor of three larger again ({\em i.e.}
$r \sim 7$)
has a transition radius that is unchanged, indicating that the smaller disc
models are sufficiently large to not affect the location of the
transition radius.

We go on to investigate the time scale  required for black-hole disc alignment.
Our results are  consistent with  the ideas  of Rees (1978), and the later work of 
Scheuer \& Feiler (1996) when the viscous diffusion coefficients acting
in and out of the plane are taken to be  equal, even though the transition radius
is much smaller in our case.
We also find that the effects of Einstein precession are to increase the 
alignment rate, as expected from a linear analysis.

Finally in section 9 we summarize our results and conclusions.

\section{Equations of motion} \label{basic-eq}
 In order to describe a compressible fluid, we adopt 
 continuity and momentum equations  in the form
\be \frac{d \rho}{dt} + \rho \nabla.{\bf v} = 0 \label{cont}\ee
and
\be \frac{d{\bf v}}{dt} = - \frac{1}{\rho} \nabla P + {\bf v} \times {\bf h}
 - \nabla \Phi + {\bf S}_{visc}
 \label{moment}\ee
where 
\be \frac{d}{dt} = \frac{\partial}{\partial t} + {\bf v}.\nabla \ee
denotes the convective derivative, $\rho$ is the density, ${\bf v}$ is the 
velocity, $P$ is the pressure,  $\Phi$ is the gravitational potential
and ${\bf S}_{visc}$ represents the viscous 
force per unit mass.  The  above equations are the standard Newtonian
equations  of fluid dynamics apart from the  addition of the
 ${\bf v} \times {\bf h}$ term  to equation (\ref{moment})
which gives  the gravomagnetic force per unit mass
experienced by the fluid in the vicinity of a rotating black hole 
to lowest post Newtonian order
(e.g. Blandford 1996).
Here ${\bf h}$ is defined by the expression
\be {\bf h} = \frac{2 {\bf S}}{R^3} - \frac{6 ({\bf S}.{\bf r}) {\bf r}}{R^5} 
\label{h} \ee
and
\be {\bf S} = \frac{ G {\bf J}}{c^2}.  \label{S} \ee
Here $R$ is the distance to the central mass and
 ${\bf r}$ denotes the coordinate vector measured from there.
It is the   gravomagnetic force  that  gives rise to Lense--Thirring precession
which   in turn leads to the alignment of   the rotation axis
 of a viscous disc  with that of the black hole.
The spin angular momentum of a Kerr black hole is given by the equation
\be {\bf J} =  \frac{a G M^2}{c}{\bf\hat{ k}}, \label{JBH} \ee
where ${\bf\hat{ k}}$ denotes the unit vector in the $z$ direction
associated with cylindrical coordinates $(r, \phi ,z)$
based on the central mass. Thus $R=\sqrt{(r^2+z^2)}.$ 
For a maximally rotating  hole, the Kerr parameter $a=1$,  with $M$  being
the mass
of the black hole, and $c$  the speed of light.

Because the effects of interest  simulated here take place many
gravitational radii from the black hole, relativistic effects are included
only in a post Newtonian approximation.

In the calculations described in subsequent sections, different prescriptions
for the gravitational force were used. In some cases, a softened Keplerian
potential was used such that the gravitational force per unit mass is
\be - \nabla \Phi = - \frac {G M}{(R^2 + b^2)^{3/2}} \bf{r} \label{Phi1} \ee
where $b$ is the gravitational softening parameter. This was adopted in order
to
prevent numerical divergences as the disc material approaches the central
object. In other cases, the effects of Einstein precession were included,
and the central object was treated as a uniform density sphere of finite size
to prevent the gravitational force from becoming infinite as $r \rightarrow 0$.
In this case the gravitational force  per unit mass outside the sphere
was taken to be
\be - \nabla \Phi = - \frac{G M }{R^3} \left( 1 + \frac{6R_+}{R}\right) {\bf r}
\label{Phi2} \ee
where $R_+= GM/c^2$, denotes the gravitational radius.
This force differs from the pseudo--Newtonian expression, often adopted 
in calculations of this kind, that gives the  correct
radius of the last stable circular orbit (e.g. Paczynski \& Wiita 1980),
but instead it gives the correct  orbital apsidal precession frequency
at large distances from the black hole.

The calculations that included softening of the gravitational force were
performed before we arrived at the idea of treating the central object as a
uniform density sphere as a means of preventing numerical divergences.
Although there is no astrophysical phenomenon that has a direct
analogy with the use of a softening parameter as described in equation 
(\ref{Phi1}), it plays a relatively small role in our calculations
of the Bardeen--Petterson transition zone. The fact that there is
an associated retrograde precession changes the behaviour slightly,
and this acts as a convenient way of illustrating the effect of the
disc rotation profile on the behavious of warps in accretion discs by comparing
the results obtained with softening with those obtained when Einstein precession
is included.

These equations are supplemented by an equation of state, which in this
case is taken to be a polytrope of index $\gamma=5/3$:
\be P =K \rho^{\gamma} \label{estate}.\ee
The sound speed is $c_s= \sqrt{dP/d\rho}.$
Energy dissipated through the action of artificial viscosity is 
simply allowed to leave the system, so that a barotropic equation of state
is assumed throughout.

\section{Precession Frequencies and the Bardeen--Petterson Effect} \label{prec-freq}

The effect of Lense--Thirring precession is  to cause
the plane of an orbit  inclined to the $(x,y)$ plane
to precess  about the angular momentum $(z)$ axis.
If a viscous disc with negligible inertia  
is set up with its midplane initially misaligned
with the $(x,y)$ plane, differential  Lense--Thirring precession will
lead to the formation of a warped disc. 
The (prograde) nodal precession frequency that arises due to the inclusion
of the gravomagnetic force term in the equation of motion (\ref{moment})
is given by
\be \omega_{z} = \frac{2 S}{R^3}, \label{prec-Lense} \ee
 where $S =|{\bf S}|.$
Note that this decreases with radius resulting in differential precession.

The Bardeen--Petterson effect is caused by the combined effects of
Lense--Thirring precession and internal disc
viscosity, which, assuming the disc has negligible inertia,
acts to align the  midplane of the inner
region of the
accretion disc with the $(x,y)$ plane.
This is because the
damping produced by viscosity, acting on the 
small scale structure (twisting up)  produced by
strong differential precession, causes the inner 
disc 
to settle into the plane of the black hole equator.

However, the outer parts of the disc remain in their original plane because the
Lense--Thirring precession rate drops off sharply as $r$ increases, such that
the internal pressure and viscous stresses  acting in the disc are able to
limit the effects of differential precession.

The radius of
transition between the two  disc midplanes is expected to occur
approximately where the rate at which the disc is twisted up by
differential precession  is balanced by the  rate at which
disc  warps are  diffused or propagated away. The rate at which the
disc is twisted up is given by $1/ \tau_{dp}=|R(d\omega_z/dR)|$,
where $\tau_{dp}$ is the differential precession time scale.
The rate at which warps
are propagated or diffused is discussed in the next section.

Important dynamical effects can also
occur because precession of orbital apsidal lines occurs
when  the central potential is non Keplerian.  The two  cases considered here
are  associated with the 
softening of the central potential  and  taking into 
account Einstein precession.

The rate of (retrograde) apsidal precession  of a near circular orbit
in the  softened 
gravitational potential (\ref{Phi1})  is given by
\be \omega_{ps} = - \frac{3}{2} \Omega \frac{b^2}{R^2}, \label{prec-soft} \ee
where $\Omega$ is  the circular orbit rotation frequency.
It can be seen that the importance of this apsidal precession  
drops off
quickly as $R >> b$. 

The rate of (prograde) Einstein  apsidal precession 
associated with the potential (\ref{Phi2})  is given by
\be \omega_{pE} = 3 \Omega \frac{R_+}{R}. \label{prec-Ein} \ee

Comparison  of these two rates of precession shows that the effects of
Einstein precession remain relatively more important further away from
the black hole than the effects of gravitational softening. 
 
Our numerical experiments show that  use of (\ref{Phi2}) rather than
just a softened potential has
 a noticeable effect on the results,
 in particular on the way in which the disc undergoes the
transition between the plane occupied by the inner disc and that occupied by the
outer disc. This is related to the effects of a non Keplerian
potential on the propagation and diffusion of warps.

\section{Misalignment and The Bardeen--Petterson Transition} \label{BP-rad}

\subsection{Small Misalignments at Large Radii}
The radius at which the disc midplane
undergoes a transition between alignment with the $(x,y)$ plane
and its initial  inclination  at large distance  can be  simply estimated
in the situation when the latter inclination is small.
To do this we consider small amplitude bending disturbances
of the disc out of the $(x,y)$ plane. These are required to have 
an inclination that is asymptotically constant
 and equal to its value at large
radii. The theory and propagation of small amplitude
bending waves has been discussed in paper  I and we may apply the analysis 
presented there.
  To make this application we first   work on the
 ${\bf v} \times {\bf h}$ term in equation (\ref{moment}).
Using (\ref{h}) this may be written
\be {\bf v} \times {\bf h} = {2S({\bf v} \times {\hat{{\bf k}}})\over R^3}
+{6S z({\bf r}\times {\bf v})\over R^5} \label{gravo}.\ee
For linear disturbances at large radii, we may regard $S$ as 
a first order quantity and  
accordingly replace the velocity
by the unperturbed Keplerian value in (\ref{gravo}).
 Thus  in cylindrical
coordinates  ${\bf v}=(0,v_{\phi},0)$ with $v_{\phi}
= \sqrt{GM/r}.$ Additionally as we are interested in a thin
disc [neglecting corrections of order $(H/r)^2$]
we replace $R$ by the cylindrical radius $r.$
Then (\ref{gravo}) becomes
\be {\bf v} \times {\bf h} =  {2S\sqrt{GM}{\hat{{\bf r}}}\over r^{7/2}}
+{6S z\sqrt{GM}{\hat{{\bf k}}}\over r^{9/2}} \label{gravo1}.\ee  
This can be considered 
[again  neglecting corrections of order $(H/r)^2$]
to arise from an effective potential
such that
\be {\bf v} \times {\bf h} = -\nabla \Phi_{gr}, \ee
with 
\be \Phi_{gr} =  {4S\sqrt{GM}\over 5 r^{5/2}}
-{3S z^2\sqrt{GM}\over r^{9/2}} \label{gravo2}.\ee   
If we now add this potential to the already existing one
we obtain the total potential governing the motion
of the fluid
\be \Phi_{Tot} \rightarrow \Phi + \Phi_{gr} .\ee
 
The governing equations are now of the same general
form as those describing bending disturbances considered in paper I.
We shall consider the low frequency limit incorporating
viscosity as in that paper. Accordingly   we adopt 
equation (19) of that paper.
This applies to disturbances
for which the $\phi$ and $t$ dependence is through a factor
$e^{i( \phi + \sigma t)}$,  $\sigma$  being the mode frequency 
and is
\be 4 g \Omega (\omega_z + \sigma){\cal I} + \frac{d}{dr} \left(
\frac{ \mu
 \Omega}{\sigma - i \alpha_1\Omega + \Omega - \kappa} \frac{dg}{dr}
\right) = 0 \label{gvisc1}\ee  
{\bf where $g = v_z'/(\Omega r)$ is the local disc tilt, $v_z'$ being the
vertical component of the perturbed velocity.}
As in paper I the symbols are defined such that
\be \kappa^2 = \frac{2 \Omega}{r} \frac{d(r^2 \Omega)}{dr} \nonumber \ee
is the square of the epicyclic frequency,         
\be {\cal I} = 
\int^{\infty}_{-\infty} \frac{\rho z^2}{c_s^2} dz, \label{I} \ee
\be \mu = \int^{\infty}_{-\infty} \rho z^2 dz. \label{mu} \ee            
Here it is supposed that the free particle
nodal and apsidal precession frequencies
$\omega_z = 2 S/r^{3},$ and $\Omega -\kappa=3\Omega R_+/r-1.5S/r^{3}$
are small compared to
the unperturbed angular velocity $\Omega$ in magnitude.

The relaxation of small amplitude bending disturbances at large radii
is described by (\ref{gvisc1}). As in paper I, this occurs
through bending waves if $\alpha_1 < H/r,$ and through
diffusion with diffusion coefficient $ {\cal D} \sim H^2\Omega /4\alpha_1$
on the characteristic time $t_{diff} =r^2/{\cal D} $ at radius $r$ 
if  $\alpha_1 > H/r.$ For $\alpha_1$ not too different from $H/r$,
as in the numerical calculations described here, the relaxation
is on the time scale required for bending waves to
propagate across the relevant length scale.
The waves propagate  with speed 
$c_B =(1/2)\sqrt{\mu/{\cal{I}}} \sim (1/2)H\Omega,$
which is  one half of a  vertically averaged sound speed.
The relaxation is  towards the solution of (\ref{gvisc1})
corresponding to the least rapid  decay in time.

For a disc of infinite extent it is possible
to look for a solution of  (\ref{gvisc1}) with $\sigma =0.$
This then takes the form
\be  g \left({\Sigma \omega_z\over \Omega }\right) + \frac{d}{dr} \left(
\frac{ c_B^2
 \Sigma}{ \Omega(- i \alpha_1\Omega + \Omega - \kappa)} \frac{dg}{dr}
\right) = 0 \label{gvisc2}.\ee    
 When $\Sigma$ and the disc aspect ratio are asymptotically
constant for $ r  \rightarrow  \infty,$ (\ref{gvisc2})
has solutions for which $g$ approaches an asymptotically constant value
$g_{\infty}$ such that 
\be g \sim  g_{\infty}  - g_1\int^{\infty}_r \frac        
{\Omega(- i \alpha_1\Omega + \Omega - \kappa)}{ c_B^2
 \Sigma} dr ,\ee
where $g_1$ is a constant which should be determined
according to an appropriate inner boundary condition.
As $r$ decreases, one eventually
reaches a point where $g$ {\em must} begin to depart
significantly from its asymptotic value $g_{\infty}$. Physically this is because
the rate at which the disc is being twisted up by the Lense--Thirring 
precession begins to dominate over the local propagation/diffusion 
of disc warp,
such that the disc is forced to lie in the equatorial plane of the black hole.
We identify the radius where $g$ has to begin to depart significantly from
$g_{\infty}$ with the Bardeen-Petterson
transition radius.  This radius can be estimated by equating it to
the length scale associated with equation (\ref{gvisc2}),
or equivalently by the more physical approach of equating the
rate at which the disc is twisted up by Lense--Thirring precession with
the rate at which warping disturbances are propagated/diffused away. 
This latter
approach is used
below.  These arguments were also  
used by Scheuer \& Feiler (1996)  to derive the transition
or warp radius (see their equation (8) and discussion).
We comment that the derivation of this
transition radius requires the disc to be  
large in comparison but not necessarily infinite. 
We emphasise that if the disc is finite but still of significantly
greater extent than the transition radius, although some
slow long term realignment and precession of the disc
will occur due to the its finite inertia, the initial relaxation
to an approximate steady state like the one considered here is expected.
This is in fact borne out by the simulation results.               

Scheuer \& Feiler (1996) gave a solution of (\ref{gvisc2})
for  a constant $\Sigma$  disc model with
constant kinematic viscosity or equivalently, assuming $\alpha_1$
 to be constant, constant
$c_B^2/\Omega.$ They adopted $\Omega =\kappa,$ however
  their solution can also be extended to the case
$(\Omega -\kappa)/\Omega =\epsilon,$ with $\epsilon$ being constant.
 The solution for $g$ (which in this context
may be identified with the complex conjugate $W^*$ of their function 
 $W$to within a sign)  is then 
\be g = g_{\infty}\exp\left(-\left({8S\Omega\over c_B^2 r}\right)^{1/2}
 \left(-\epsilon+i\alpha_1 \right)^{1/2} \right)\label{SFg}
 .\ee            
As long as $\alpha_1 \ne 0,$ this solution is such that 
$g \rightarrow 0$, for $r \rightarrow 0.$

However,  for $\epsilon >0,$
corresponding to prograde precession, as in the Einstein case,  and 
 small  $|\alpha_1/\epsilon|,$ the solution
  exhibits many   wavelike oscillations which may
be interpreted as {\it inward} propagating bending waves
excited in the transition region. This is in contrast
to the situation of retrograde precession, $\epsilon < 0,$ 
as in the softened case, where
the waves are evanescent.
This phenomenon has been already  noted by Ivanov \& Illarianov (1997)
and,  as we shall  see,
it has some consequences for the calculation of the
torque exerted between the black hole and disc in the limit 
of small $\alpha_1.$ But note that solutions with many
oscillations in the inner disc
 appear in a linear theory. It is likely that non linear effects
 are likely to make very short wavelength waves disappear.

The case of Einstein precession has $\epsilon = 3R+/r,$ which is not
constant. However a solution of the
Scheuer \& Feiler (1996) type may be found
in this case  for constant $\Sigma$ in the limit of  vanishing
$\alpha_1,$  if we assume the kinematic
viscosity or equivalently $c_B^2/\Omega \propto 1/r.$
This is

\be g = g_{\infty}\exp
\left(-i\left({24S R_+\Omega\over c_B^2 r}\right)^{1/2}
 r^{-1/2}\right) \label{SFg1}
 .\ee

\section{The Transition Radius} \label{Rtrans}
In order to obtain an estimate of the transition radius for the warp,
we need to equate
the rate at which the disc is twisted up by the rate at which warping
disturbances are propagated across the disc. We note that estimates
of the transition radius based on this approach should 
provide results that are accurate to within a factor of order unity.
The physical processes of twisting the disc up and propagating/diffusing
 the warp
occur over a range in radii such that the transition between aligned
and misaligned planes will also occur over a range in radii,
rather than at a single location. 
In the diffusive limit where
$H/r < \alpha_1,$ and $\alpha_1 > |\Omega -\kappa|/\Omega,$  we
equate the rate of twisting up by differential precession
and the rate of diffusion of warps.
The rate at which the
disc is twisted up is given by $1/\tau_{dp} = |r(d\omega_z/dr)| =6S/r^3.$ 
In the diffusion limit,
the warping of the
disc is counteracted by  diffusion of the warp, which acts  at
a rate  $1/t_{diff}$ where $t_{diff}=r^2/{\cal D}$ and
${\cal D} = c_B^2/(4 \alpha_1 \Omega)$ is the diffusion coefficient.
Equating these two rates leads to an expression
from which the transition radius can be found:

\be R_{T1} = \left\{ \frac{24 \alpha_1 S}{\sqrt{G M}} \left( \frac{H}{r} \right)^{-2}
\right\}^{2/3} \label{r_t1} \ee
This should be contrasted with the expression for the transition radius,
$R_{BP}$,
obtained using thin disc theory in which the effects of pressure forces
are neglected and an isotropic viscosity is assumed (Bardeen \& Petterson 1975;
Hatchet, Begelman, \& Sarazin 1981):
\be R_{BP} = \left\{ \frac{6 S}{\alpha \sqrt{G M}} 
\left( \frac{H}{r} \right)^{-2} \right\}^{2/3}. \label{r_bp} \ee
Here $\alpha$ is the standard  viscosity parameter associated with
the radial accretion. 
It is apparent that if $\alpha << 1$, then $R_{BP} >> R_{T1}$ and the
transition radius predicted by (\ref{r_bp})
will lie much further away from the black
hole, with the warp extending over a much greater proportion of
the accretion disc than when (\ref{r_t1}) is used. 
This arises because the viscous damping of the
resonantly driven horizontal motions
included  when (\ref{r_t1}) is used
leads to a larger diffusion coefficient,
and thus to a more rapid damping of the warp.

It should be noted that $R_{T1}$ given by (\ref{r_t1}) continues to decrease
as $\alpha_1$ decreases. Here we remark that for a strictly
Keplerian disc the diffusive behaviour
of warps used in (\ref{r_t1}) requires for its validity
that $\alpha_1 > H/r.$ When $\alpha_1 \sim H/r,$ warps
are diffused across the disc on essentially  the sound crossing time.
At the beginning of the transition to the 
wave communication regime,  we replace $\alpha_1$ 
by $H/r$ in (\ref{r_t1}), giving a second estimate of the transition radius
\be R_{T2} = \left\{ \frac{24  S}{\sqrt{G M}} \left( \frac{H}{r} \right)^{-1}
\right\}^{2/3}. \label{r_t2} \ee

Note further that the disc is not strictly Keplerian in which case
apsidal precession, which inhibits
the communication of warps, should be taken into account. This modifies
(\ref{r_t1}) for the diffusive regime to provide a third estimate of the 
transition radius.
\be R_{T3} = \left\{ \frac{24 S \sqrt{(\alpha_1^2\Omega^2+(\Omega-\kappa)^2)}
 }{\Omega\sqrt{G M}} \left( \frac{H}{r} \right)^{-2}
\right\}^{2/3} \label{r_t3}.\ee
Note that if we take account of the fact that $\Omega \ne \kappa,$
because $S \ne 0,$
but ignore the effect of Einstein precession, then we 
recover (\ref{r_t2}) 
in the limit of small $\alpha_1.$

The  above discussion suggests that to estimate the Bardeen-Petterson radius
we should take the maximum value found from using (\ref{r_t1}), (\ref{r_t2}) or
(\ref{r_t3}). In this way unrealistic efficiency in warp communication
is avoided.
However, we remark that the transition radius is  
expected to be many gravitational
radii from the black hole justifying the post Newtonian approximations
in estimating its location. We further remark that the appropriate
values of $\alpha_1$ for our disc models were calibrated using
calculations similar to those presented in paper I and those in
Nelson, Papaloizou, \&
Terquem (1999). These values are given in the introduction.

\section{Numerical Method} \label{Num}
The set of fluid equations described in section (\ref{basic-eq})
are solved using smoothed particle hydrodynamics
(Lucy 1977; Gingold \& Monaghan 1977). The SPH code used to perform the
calculations presented in this paper is identical to that used in paper I,
and we refer readers to this paper and references therein
for a description of our numerical scheme.

\section{Calculations of the Bardeen--Petterson Effect} \label{Bard-calc}
We have performed a number of calculations to examine the structure
of an accretion disc produced when in orbit about a Kerr black hole,
with the effects of Lense--Thirring precession included in the equations
of motion. The calculations assume that the angular momentum
vectors of the disc material and the spinning black hole are
initially misaligned by some amount, with the system being evolved until a
quasi steady state is achieved.
The main issue that we wish to address in these simulations
is the location of the so--called `Bardeen--Petterson radius'.
This is the region of transition in the disc between the inner part,
which in a steady state is expected to lie in the equatorial plane of
the black hole due the combined effects of differential precession and 
viscosity, and the external part which is expected to remain in its original 
plane if the disc is effectively infinite. Due to the limitations of
computational tractability, we are only able to consider discs which
are of finite size. We do not expect this to affect the results on the 
transition radius, provided that the outer disc radius is sufficiently far from
the transition zone, as is the case in our disc models.
The initial conditions of each of the calculations were varied,
such that the effects of changing the disc thickness, the exact form
of the central potential, the number of particles, the disc radius, and the
inclination angle between the black hole equator and  initial disc midplane   could be 
studied. These initial conditions are described in detail in the following
section, which is followed by a description of the results from the
numerical calculations.

\subsection{Initial Conditions} \label{BP-init}
In each of the Bardeen--Petterson calculations, the
central mass $M=1$, the gravitational constant $G=1$, $a=1,$ and ${\bf S}=0.008$
in equation (\ref{h}). This 
leads to a value of the gravitational radius of $R_+=0.04$ in our computational
units. The calculations can be divided essentially
into two groups, those in which the central Keplerian potential
was softened, and those in which no softening was used but in which
the effects of Einstein precession were included.
We will describe the former group of calculations first, followed by the
latter group. One additional calculation was performed in which the effects of
Einstein precession were neglected and no softening of the central potential 
was used,
leading to a disc with a Keplerian rotation profile. 
All calculations are presented in table~\ref{tab1} , with the predicted value
of $R_{T2}$ from equation (\ref{r_t2}) being shown in the eighth column, and
that measured from the SPH calculations, $R_{TS}$ being presented in the last.
In this paper we define $R_{TS}$,
the transition radius measured from the simulations,
to be the radius at which the disc tilt has a value that is half way between
that of the inner and outer parts of the relaxed warp disc models, 
unless otherwise
stated.

The gravitational softening parameter used in equation (\ref{Phi1})
was $b=0.2$. A number of disc models were employed to examine the
Bardeen--Petterson effect.
An initial calculation used a disc model of radius $R=1$ and
employed $N=20,000$ particles. This calculation had a midplane
Mach number of ${\cal M} \simeq 12$.  Two calculations
were performed with disc models of radius $R=2$ and with $N=52,000$.
These calculations had ${\cal M} \simeq 12$ and $30$ respectively.
Two calculations were performed with a disc of radius $R=\sqrt{2}$ and
$N=102,000$ particles, one with ${\cal M} \simeq 9$, and the other with
${\cal M} \simeq 12$. An additional calculation was performed with
$R=\sqrt{2}$, $N=60,000$, and ${\cal M} \simeq 12$. In all cases, the
inclination between the black hole equator and the initial disc midplane
was $i=10^o$.

The calculations in which no gravitational softening was employed,
and which included the effect of Einstein precession, had the central
black hole treated as a uniform density sphere of finite size ($b=0.2$)
in order
to prevent numerical divergences as the disc material extends towards
$r \rightarrow 0$. 
Six calculations of this type were computed overall. 
Five of them had $R=2$ and $N=52,000$, two of which had
had $i=10^o$ and ${\cal M} \simeq 12$ and $30$, two of which
had $i=30^o$ with ${\cal M} \simeq 12$ and $30$, and one had $i=10^o$ and 
${\cal M}=5$. The sixth calculation had $i=10^o$, $R=7$, $N=200,000$, and
${\cal M} \simeq 14$.
An additional calculation was performed in which the effects of 
Einstein precession were neglected and no softening of the gravitational
potential was used. This calculation had $i =30^o$, ${\cal M}=12$, $R=2$, and
$N=52,000$, and was performed to provide a comparison case with the other
high inclination runs.

All calculations were initiated  with the disc  
orbiting  the black hole, and with the inclination between the
black hole equator and the {\em original} disc plane being constant
throughout the calculations. The calculations were all evolved towards
a quasi steady state, such that a precessing, but otherwise
almost stationary warped structure was obtained in each case.
The results of these calculations are described in the following section.

 \begin{table*}
 \begin{center}
 \begin{tabular}{llllllllllllr}  \hline \hline
 Run &  $i$ & $R$ & ${\cal M}$ & $ b_{soft}$
   & Ein  & $N$ &
$R_{T2}/R_+$  & $R_{TS}/R_+$ & $t_{run}/\tau_{dp}$ \\
&       &    & & & Prec?    & $(\times 10^3)$ &             &  &        \\ \hline

&       &    & & &    &    &          &                &         \\
S1 & 10   & 1 & 12 & 0.2 & No & 20 & 43 & 17 & 2.4 \\
&       &    & & &    &    &                  &        &         \\
S2 & 10   & 2 & 12 & 0.2 & No & 52 & 43 & 16 & 2.4 \\
&       &    & & &    &    &                 &         &         \\
S3 & 10   & $\sqrt{2}$ & 12 & 0.2 & No & 60 & 43 & 16 & 2.1 \\
&       &    & & &    &    &                    &      &         \\
S4 & 10   & $\sqrt{2}$ & 12 & 0.2 & No & 102 & 43 & 16 & 2.4 \\
&       &    & & &    &    &                     &     &         \\
S5 & 10   & $\sqrt{2}$ & 9 & 0.2 & No & 102 & 33 & 10 & 14 \\
&       &    & & &    &    &                          & &         \\
S6 & 10   & 2 & 30 & 0.2 & No & 52 & 80 & 33 & 2 \\
&       &    & & &    &    &                  &        &         \\
E1 & 10   & 2 & 12 & 0 & Yes & 52 & 43 & 14 & 2.6 \\
&       &    & & &    &    &                 &         &         \\
E2 & 10   & 2 & 30 & 0 & Yes & 52 & 80 & 28 & 2.9 \\
&       &    & & &    &    &                 &         &         \\
E3 & 30   & 2 & 12 & 0 & Yes & 52 & 43 & 20 & 5 \\
&       &    & & &    &    &                 &         &         \\
E4 & 30   & 2 & 30 & 0 & Yes & 52 & 80 & 25 & 5 \\
&       &    & & &    &    &                 &         &         \\
E5 & 10   & 2 & 5 & 0 & Yes & 52 & 24 & 0 & - \\
&       &    & & &    &    &               &           &         \\
E6 & 10   & 7 & 14 & 0 & Yes & 200 & 48 & 19 & 2.4\\
&       &    & & &    &    &                &          &         \\
K1 & 30   & 2 & 12 & 0 & No & 52 & 43 & 23 & 5 \\
&       &    & & &    &    &                &           &         \\
 \hline \hline
 \end{tabular}
 \end{center}
 \caption{\label{tab1}The first column provides the label 
 given to each run, the 
 second column provides the initial inclination. The third column gives the 
 disc radius, and the fourth column gives the midplane Mach number of the 
 disc model. The fifth column gives the gravitational softening length,
 the sixth column indicates whether or not Einstein precession was included,
 and the seventh column gives the number of particles used.
 The eighth column gives the predicted value of the transition
 radius from equation (\ref{r_t2}), and the ninth column gives the value of
 the transition radius measured from the simulations, in units of the
 gravitational radius. The 
 last column provides the time over which the simulations were run 
 in terms of the differential precession time at the outer edge 
 of the transition zone.}
 \end{table*}

\section{Results}
In this section we present and discuss the results of the calculations described
in section (\ref{BP-init}). The results of the calculations that were performed
using a softened gravitational potential are described first, followed by
the calculations in which the effects of Einstein precession were included.

\begin{figure*}
\epsfig{file=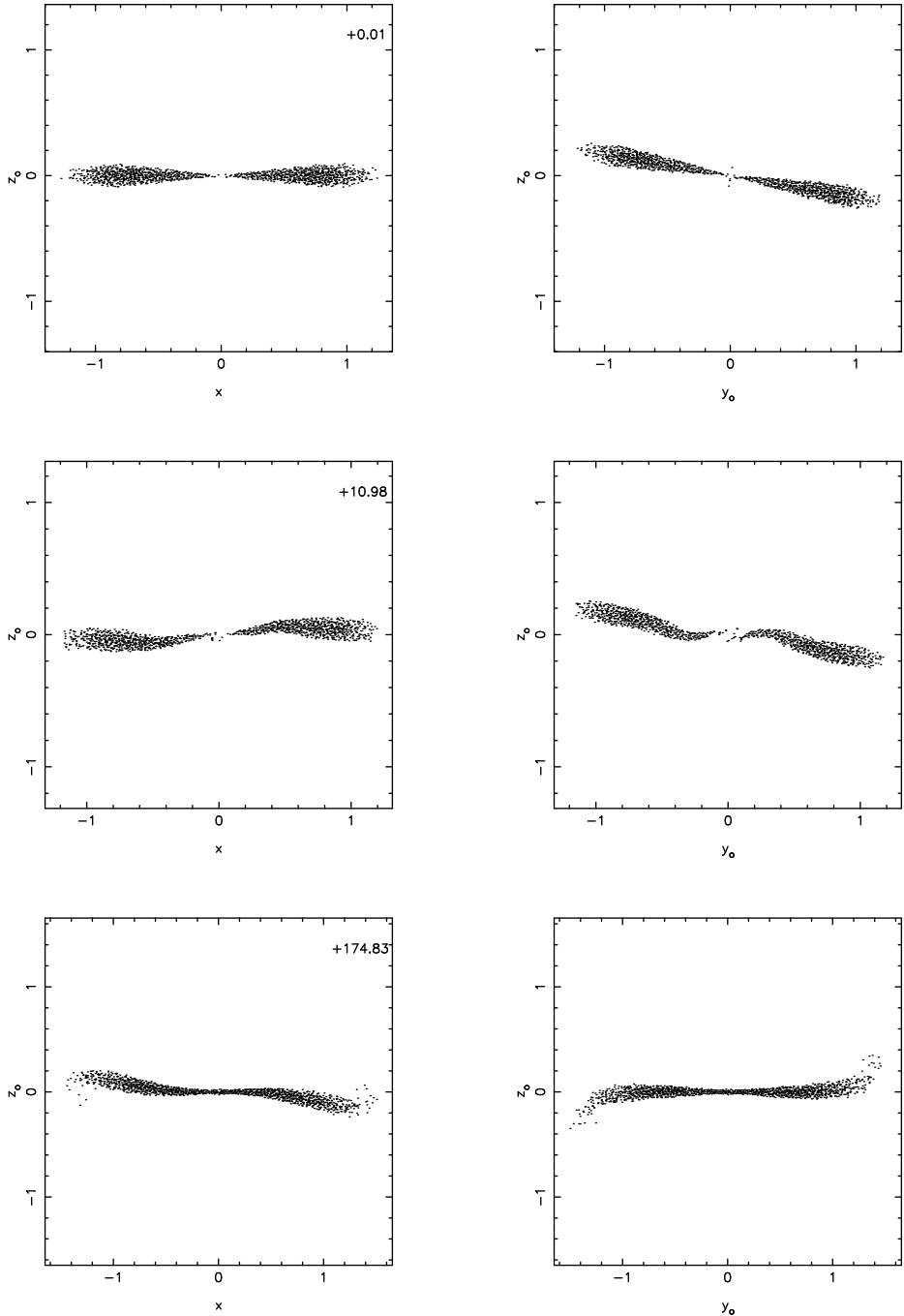, width=14cm}
\caption{This figure shows particle projections of cross sectio
nal slices
through the disc in calculation S1. The time is shown in units of
$\Omega (R=1)^{-1}$ in the top right corner of each left hand panel.}
\end{figure*}

In the discussion of the results that follows, we will be describing
the evolution of the angle of inclination between angular momentum vector
of the black hole, ${\bf J}_{bh}$, and the specific angular momentum vector of
the disc material, ${\bf J}_d$. This quantity is defined through the
expression
\be i(r) = \arccos{\left(\frac{{\bf J}_{bh} \; . \; {\bf J}_d(r)}{ | {\bf J}_{bh} | 
|{\bf J}_d(r) |} \right)},
\label{i} \ee
and may be a function of radius through the disc, such that the disc is warped.
The calculations are initiated with this quantity being constant throughout the
disc. We note that $g$, introduced in equation (\ref{gvisc1}), and $i$ are
equivalent.
We will also be describing the precession of the local angular momentum
vector in the disc, projected onto the fixed $x$--$y$ plane of the
black hole equator. This precession angle is defined by the expression
\be  \beta(r) = 
\arccos{ \left(\frac{{\bf J}_{bh} \times 
{\bf J}_d(r)}{| {\bf J}_{bh} \times {\bf J}_d(r) | } \; . \; \hat{{\bf y}_o} \right)}, \label{beta} \ee
and may vary as a function of radius in the disc, such that the disc is
twisted. Here, $\hat{{\bf y}_o}$ is the unit vector pointing in the $y$ direction
located in the fixed equatorial plane of the black hole.
The calculations are initiated with $\beta= \pi/2$ throughout the disc.

\begin{figure}
\epsfig{file=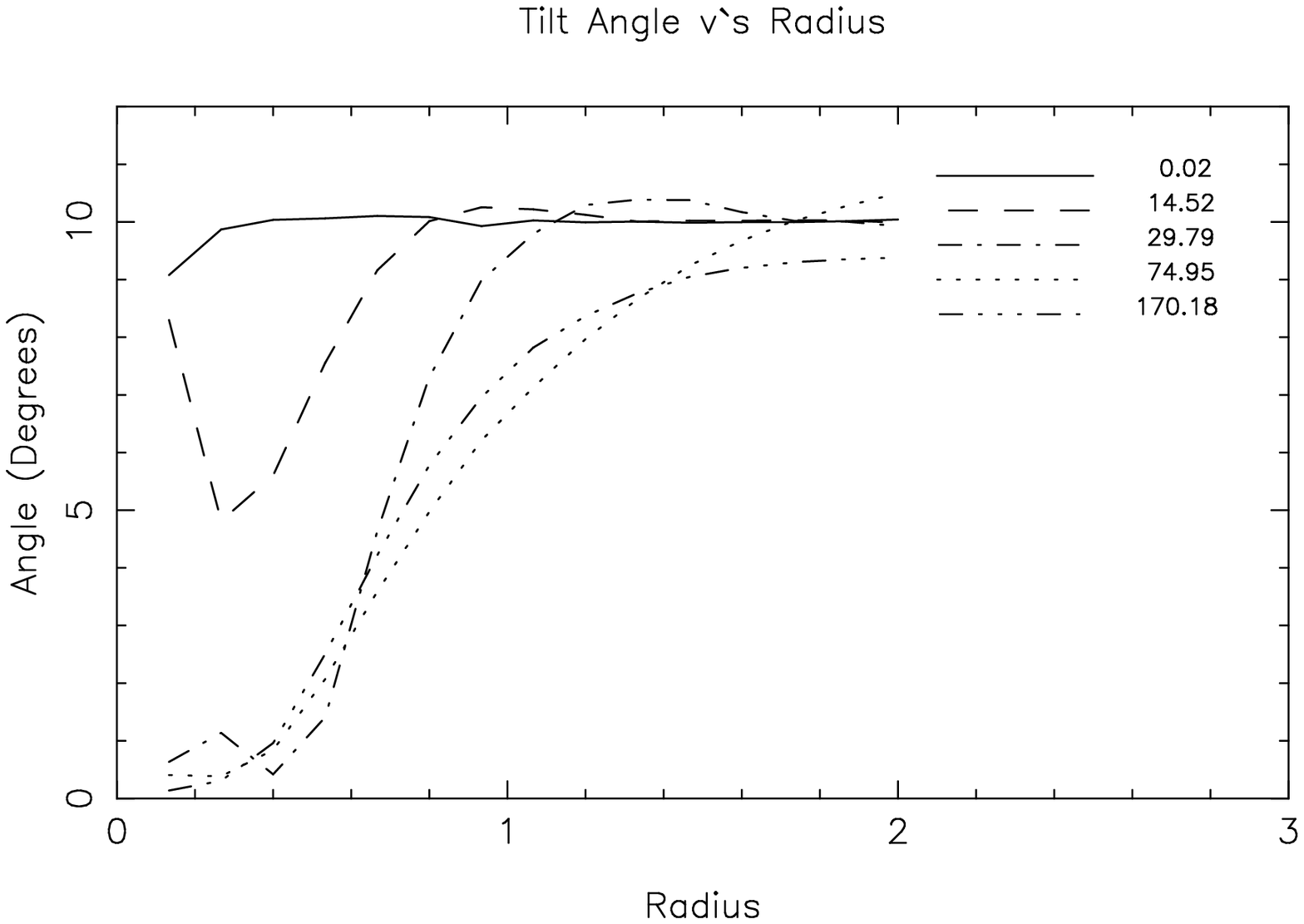, width=\columnwidth}
\caption{This figure shows the evolution of disc tilt as a function
of radius for the calculation S2, with midplane mach number
${\cal M} \simeq 12$.}
\end{figure}

The results shown in the particle projection plots are presented in a 
fixed reference frame centred on the black hole. The initial inclination 
between the disc midplane and the black hole equator is provided by rigidly
rotating 
the disc about a diameter coincident with the $x$ axis in this fixed reference
frame.

\begin{figure}
\epsfig{file=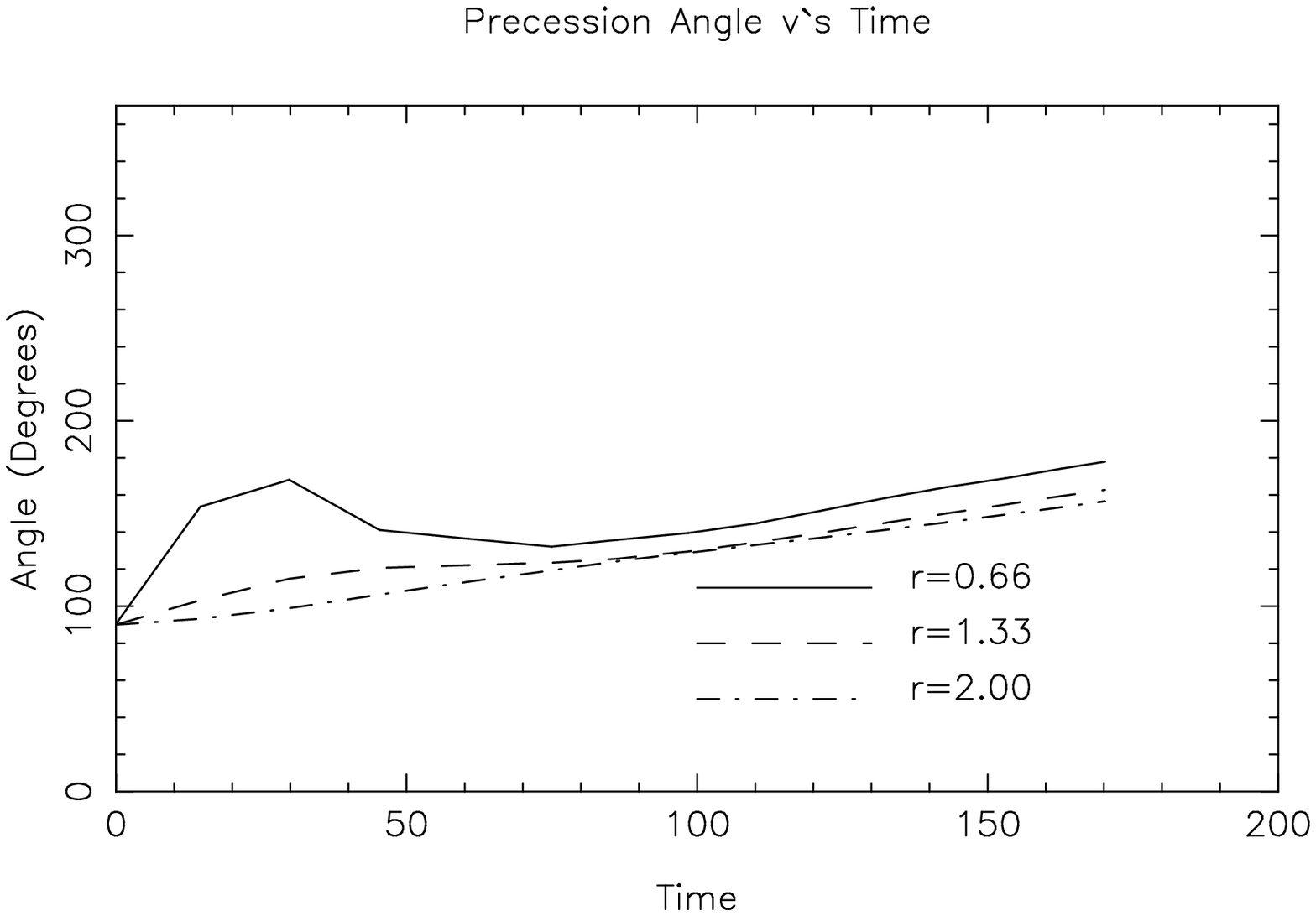, width=\columnwidth}
\caption{This figure shows the precession angle as a function of time
for three different disc annuli in calculation S2. Note the initial
differential precession which relaxes towards near solid body precession
after $t \simeq 100$. The outer radius of each annulus is indicated in the
figure panel.}
\end{figure}

The evolution of all of the disc models in the simulations is as follows.
Initially, there is a short period of rapid evolution during which the
inner parts of the disc undergo strong differential precession and are
forced to lie in the symmetry plane of the black hole out to a transition
region, beyond which the discs remain in their original plane.
During this time,
warping disturbances propagate through the disc either as bending waves or
through diffusion, and a warped disc configuration is set up on the time scale
required for these warps to propagate across the disc.
This time scale for the the warped configuration and the transition zone to be
set up also corresponds to the differential precession time scale
measured at the transition radius, $\tau_{dp}$. The discs were all evolved for
a time $>> \tau_{dp}(R_{TS})$ to ensure that the position of the transition
zone had been firmly established. The run time of each calculation, in terms
of the relevant differential precession time $\tau_{dp}$, is given
in table~\ref{tab1}.
There then follows a period of evolution that occurs on a much
longer time scale involving near--rigid body precession of the disc, and
a slow alignment of the outer parts of the disc with the symmetry plane
of the central black hole. This period of evolution arises because
of viscous coupling between the misaligned inner and outer parts of the disc,
and as a result of residual torquing by the black hole beyond the transition
radius. It does not arise because warping disturbances are being propagated
to large radii from within the transition zone of the disc.
The communication of warps by bending waves or diffusion appears to be
confined within the transition zone, exactly as expected from the
linear theory. This is evidenced by the fact that the
outer regions evolve much
more slowly than they would if they were being torqued into
alignment as a result of warps propagating from the inner regions of the disc.
We have examined the outer parts of the discs in our models, and find no
evidence of warping disturbances reaching them from their 
central parts over time scales that are typically $\sim$ few 
hundred $\Omega(R=1)^{-1}$. We interpret this to mean that using disc
models with larger outer radii will not result in changes to the 
locations of the transition radii
that we find in our models. This is confirmed by run E6.

\subsection{Calculations Employing Gravitational Softening}
A number of calculations were performed in which the central gravitational
potential was softened and the effects of Einstein precession were neglected.
These calculations were performed with ${\cal M} \simeq 9$, 12 or 
$30$ ({\em i.e.} with $H/r \simeq 0.11$, 0.08 or 0.03).

\begin{figure}
\epsfig{file=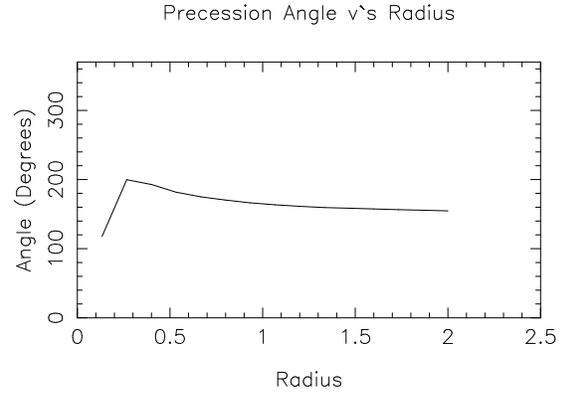, width=\columnwidth}
\caption{This figure shows the degree of twist in the disc at the end of
calculation S2. Note that the inner disc is twisted relative to the
outer disc by an angle of $\sim 25$ degrees.}
\end{figure}

\subsubsection{${\cal M} \simeq 12$ Calculations}

A time sequence showing the evolution of calculation S1, listed in 
table~\ref{tab1}, 
is shown in Fig. (1). Projections of particles contained within
a thin slice centred about the $x$ axis (left panels) and the $y$ 
axis (right panels) are
presented, with the time corresponding to the panels being shown in units of
$\Omega^{-1}$ at $R=1$ in the top right hand corner of the left hand panels.
The top two panels show the disc at the beginning of the calculation, with the initial
tilt of the disc relative to the fixed coordinate system being apparent in the right hand panel.
As the calculation proceeds, the Lense--Thirring precession twists up the
inner parts of the disc, which are forced eventually to lie in the
equatorial plane of the black hole, with the outer disc remaining close to its
original plane. The middle two panels show the disc at an intermediate state
of its evolution, during the time in which
the inner parts of the disc are still in the process of
flattening out into the equatorial plane of the black hole.
The final two panels show the disc after it has evolved towards a quasi--steady state configuration, in which a smoothly varying warped disc has been established
which undergoes approximate solid--body precession. We comment that
this non zero precession rate is a consequence of having a finite disc.
If the disc were to extend to arbitrary large radii, the angular
momentum content may tend to arbitrary large values while the total mass
remained finite.  Then the precession period would approach arbitrary large values.

\begin{figure}
\epsfig{file=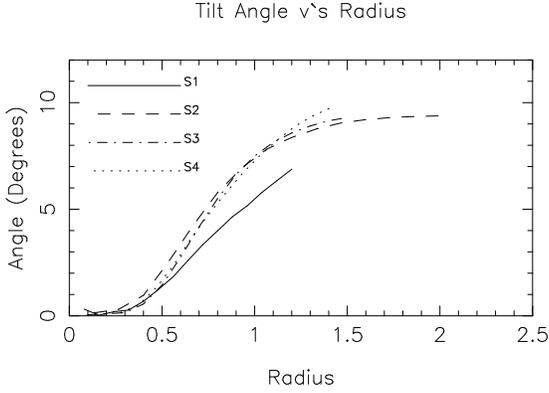, width=\columnwidth}
\caption{This figure presents a comparison of tilt profiles
obtained in calculations
S1, S2, S3, and S4 which each had midplane Mach number of ${\cal M} \simeq 12$.
Note that each calculation produces a  fully aligned region of the disc
that extends out to $R \simeq 0.4 = 10 R_+$.}
\end{figure}

Fig. (2) shows the time evolution of the disc tilt
versus radius for the calculation S2 shown in table~\ref{tab1}, with
the unit of time being $\Omega^{-1}$ evaluated at $R=1$.
The angular momentum vector of the disc
at each radius is initially tilted by $i = 10^o$ with respect to the angular
momentum vector of the black hole. As the calculation proceeds the inner parts
of the disc begin to lie in the equatorial plane of the black hole, and eventually
a disc configuration is developed in which the disc tilt varies smoothly as
a function of radius, and with the radius of transition between the
inner and outer planes being at around $R_{TS} \sim 0.6$, which is equivalent
to $R_{TS} = 15 \; R_+$ in our units. To recap, we define the
transition radius measured in the simulations, $R_{TS}$, 
to be the radius at which the disc angle of
tilt is half way between that of the outer part of the disc and the inner
part of the disc. 
In the case of calculation S2, the transition zone approximately occupies the 
range of radii 0.4 -- 1.5. The differential precession time measured
at $R=1.5$ is $\tau_{dp}(1.5)=70 \; \Omega(R=1)^{-1}$. The calculation
is evolved for 170 $\Omega^{-1}$ without any apparent change in the transition
radius between the times 100 -- 170 $\Omega^{-1}$.
We find in this calculation, and in 
the others also, that the position of $R_{TS}$ agrees reasonably well with
the semi--analytic formulae presented in section {\ref{BP-rad}, such as
equation (\ref{r_t2}), to within a factor of 2 or 3, thus justifying the
physical arguments leading to these expressions. We note that
the choice of defining
$R_{TS}$ to be the radius corresponding to the half way point of the 
warped region is somewhat arbitrary, and that a definition of $R_{TS}$
being the position where the inclination approaches its value at the
outer disc edge would produce values of $R_{TS}$ in table~\ref{tab1} that are
in greater agreement with the tabulated values of $R_{T2}$.
We also comment that the inclination of the outer disc boundary is free,
and thus its location should not affect the position of the transition radius
if it is sufficiently distant (compare models S1, S2, S3, S4 in Fig.[5],
and E6 in Fig. (\ref{fig13}).
The equations (\ref{r_t1}), (\ref{r_t2}), and (\ref{r_t3}) that predict
the transition region were derived
under the assumption of linearity of the warp. We note that the warp in the Mach
30 cases is nonlinear, as described in paper I, such that we should expect 
a poorer agreement between the transition radii measured in the simulations
and those predicted using equations such as (\ref{r_t2}).

\begin{figure}
\epsfig{file=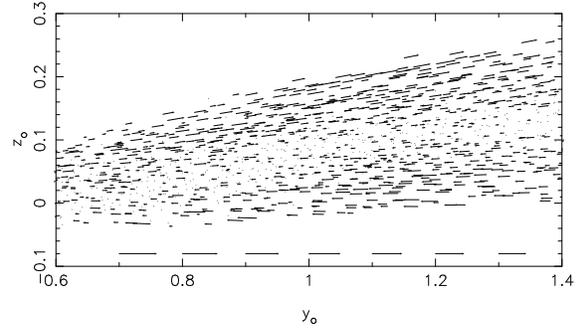, width=\columnwidth}
\caption{This figure shows the vertical shearing motion generated in the
warped accretion disc modeled in calculation S4. This shearing motion is
observed in all models in the regions of the disc where the curvature,
$\partial g / \partial r$, is
non zero.}
\end{figure}

Fig. (3) shows the time evolution of the nodal precession angle, $\beta$, for
different annuli of the accretion disc, computed from the results of 
calculation
S2. The definition of $\beta$ is given by equation (\ref{beta}), such that at 
time $t=0$, $\beta=90^o$ for the whole disc. In Fig. (3), the disc has 
been subdivided into three annuli which each lie in the 
range $0 \le r < 0.66$, 
$0.66 \le r < 1.33$, and $1.33 \le r \le 2$, respectively.
As the calculation proceeds, it may be observed that the disc initially 
undergoes a period of differential precession, with the inner disc precessing
more rapidly than the outer disc, before settling down to a quasi steady state
in which the disc precesses approximately as a rigid body.
The degree of twist in the disc at the end of the calculation is represented in
Fig. (4). This figure indicates that the angle between the line of nodes
of disc annuli in the inner part of the disc, say at $r=0.6$, and the outer disc
edge is $\sim 25^o$, showing that the disc has developed a $ \sim 25^o$ twist
before reaching a quasi equilibrium state.

A comparison between each of the runs with midplane Mach numbers ${\cal M} \sim
12$ is presented in Fig. (5). This figure shows the results from each of the
runs S1, S2, S3, and S4 presented in table~\ref{tab1} . It is apparent that the
radial variations in tilt, and in particular the positions of the
transition radii, are similar in each case, even though the resolution of each
calculation differs, indicating that convergence of the results has 
been attained. We note that the position of 
outer radius of the disc does not seem
to alter the location of the transition region, since the outer disc lies 
sufficiently beyond the warped region.
The curve corresponding to the R=1 disc (calculation S1) shows that
the outer parts of this disc are tending towards alignment with the
black hole equatorial plane. This is expected for all discs of finite
extent, and hence finite angular momentum and moment of inertia, since there
is a torque between the inner and outer parts of the disc that attempts to 
bring them into alignment. The rate at which this occurs obviously depends on
the angular momentum content, and hence radius, of the disc, with a smaller
disc being aligned more quickly. The larger (R = $\sqrt{2}$ and $R=2$) discs 
will also eventually evolve towards alignment, but on a time scale that is
longer than the time for which these calculations were run. This issue 
will be discussed in greater detail in section \ref{align}.

As described in paper I, the result of warping a 
vertically stratified, gaseous accretion disc is to set
up vertical shearing motions, such that the perturbed radial velocities, $v_r$,
are an odd function of $z$. The warped discs generated by the action of
Lense--Thirring precession indeed show such a kinematic feature over 
those portions of the disc where the inclination, $i$, varies with radius.
This fact is illustrated by Fig. (6), in which the radial velocities
of the particles contained within a thin slice of the disc centred about the 
$y$ axis are represented by velocity vectors. The arrows located at the bottom
of the figure indicate the magnitude of the midplane sound speed at each 
position. The data used to generate this figure were obtained from the 
calculation S4.

\begin{figure*}
\epsfig{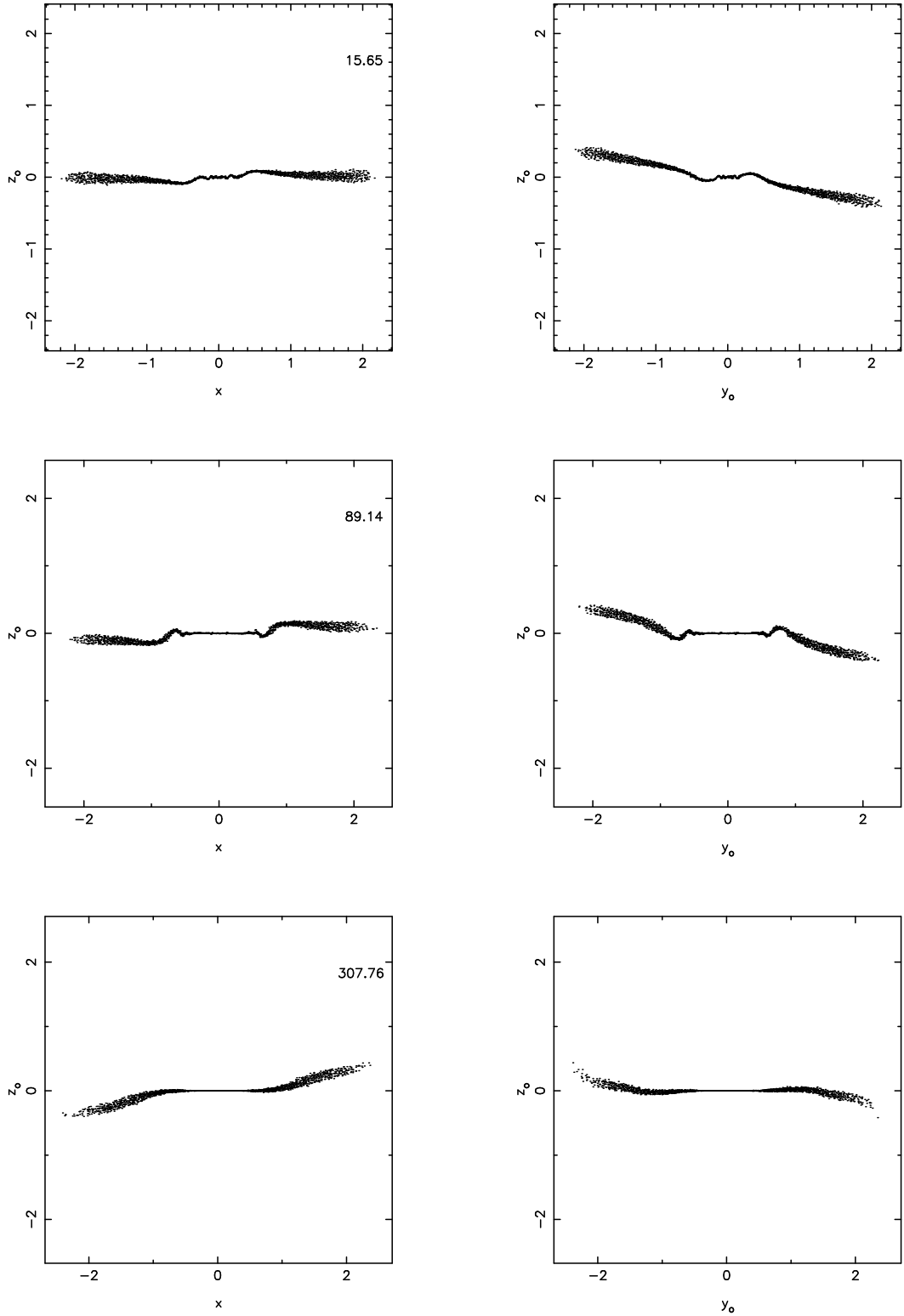}
\caption{This figure shows particle projections of cross sectional slices
through the disc in calculation S6. The time is shown in units of
$\Omega(R=1)^{-1}$ in the top right corner of each left hand panel.
Note that the `wiggles' observed in the disc are transient, and damp out by
the end of the calculation.}
\end{figure*}

\subsubsection{Changing the Disc Thickness -- ${\cal M} \simeq 30$ and ${\cal M} \simeq 9$ Calculations}

A time sequence showing the evolution of the disc during the calculation
S6 is shown in Fig. (7). Projections of particles contained within a slice
centred about the $x$ axis (left panels) and the $y$ axis (right panels)
are presented. The time corresponding to each panel is shown
in the top right hand corner of each left hand panel in units of $\Omega^{-1}$ 
evaluated at $R=1$. The first two panels show the disc after it has
evolved for $t \simeq 15 \; \Omega^{-1}$, and the existence of apparent
`wiggles' may be observed in the inner parts of the disc. These features
are transient phenomena and are caused by the differential 
precession of the inner disc. As time proceeds, the wiggles damp away, as 
the inner disc flattens into the equatorial plane of the black hole, 
eventually leaving a disc
with a warp that is slowly varying as a function of radius.
We notice that the inner region of the disc that becomes aligned with the black hole equator is of much greater radial extent that that occurring in Fig. (1),
as expected from the discussion presented in section \ref{BP-rad}.

\begin{figure}
\epsfig{file=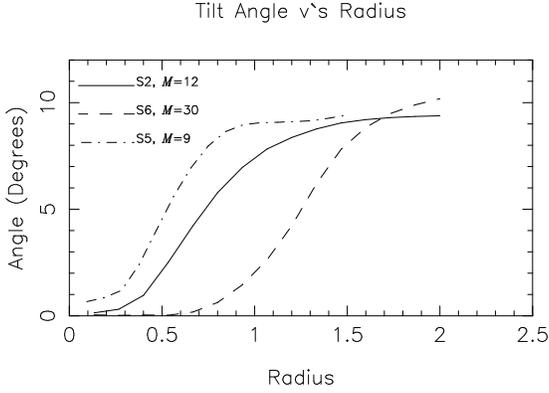, width=\columnwidth}
\caption{This figure shows the variation of disc tilt with radius at the end
of the calculations S2, S5, and S6. The value of the Mach number at 
the disc midplane is given in the figure panel alongside the figure labels.
Note that the transition radius $R_{TS}$
moves in as
the midplane Mach number is decreased.}
\end{figure}

The variation of disc tilt as a function of radius is plotted for three
different runs in Fig. (8) in order to examine the effect of changing the disc
thickness. The midplane Mach number of each run is
${\cal M} \sim 9$ ({\em dot--dashed} line), ${\cal M} \sim 12$ 
({\em solid} line), and ${\cal M} \sim 30$ ({\em dashed} line), and 
it is apparent that the transition radius increases
with increasing ${\cal M}$. This result is expected from the 
analytic estimates of $R_{T1}$ and $R_{T2}$
 presented in section \ref{BP-rad}, since
a thicker disc diffuses disc warp more efficiently.
The value of $R_{TS}$ changes from $R_{TS} \sim 10 R_+$  to $16 R_+$, and $33 R_+$
as ${\cal M}$ changes from ${\cal M} \sim 9$, 12, and 30, respectively.
It should be noted that these values of $R_{TS}$ are very much smaller than 
those predicted using the thin viscous disc theory 
neglecting hydrodynamical effects, (Bardeen \& Petterson 1975),
and confirm the previous assertions of Papaloizou \& Pringle (1983), 
and Kumar \&
Pringle (1985), that taking into account the full disc hydrodynamics results 
in
a transition radius that lies much closer to the central black hole.

\begin{figure}
\epsfig{file=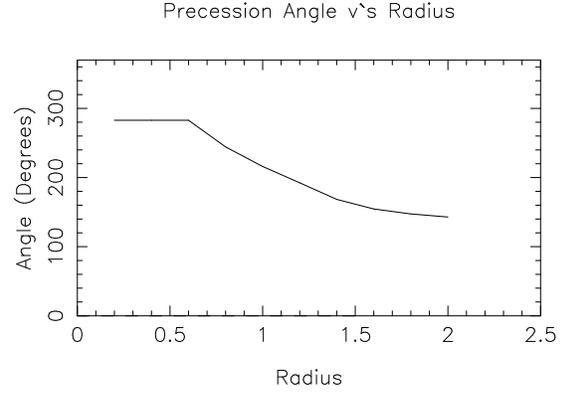, width=\columnwidth}
\caption{This figure shows the degree of twist in the disc at the end of
calculation S6, which has a midplane Mach number of ${\cal M} \simeq 30$.
Note that the inner disc is twisted relative to the
outer disc by an angle of $\sim 90$ degrees, which should be contrasted with
that in calculation S2, where ${\cal M} \simeq 12$.}
\end{figure}

The degree of twist in the ${\cal M} \sim 30$ disc is shown in Fig. (9), and
indicates that the twist angle between the line of nodes in the inner
($R \sim 0.9$) and outer parts ($R=2$) of the disc is $\sim 90^o$, which is
significantly larger than was found to be the case for the ${\cal M} \sim 12$
disc. This larger twist occurs because communication through the disc by 
bending waves/diffusion is much less efficient in this case, so that the disc
needs to acquire a more distorted configuration before it can 
obtain a quasi steady state structure which undergoes approximate solid body 
precession.

We note that although the numerical resolution in the ${\cal M} \sim 30$ 
calculations is reduced relative to the ${\cal M} \sim 12$ calculations,
a plot of the velocity field similar to that shown in Fig. (6) produces 
a similar picture of a disc in which the vertical shearing motion
extends over those regions of the disc in which $i$, 
the disc inclination, varies with
radius ({\em i.e.} in those regions where the curvature of the disc is 
non--zero). We also found that these motions
were subsonic which is consistent with the idea that
damping due to shocks keeps them down to this level.
This indicates that the thinner disc models still have
their vertical structure modeled to a reasonable degree of accuracy, such that
it is able to produce this
vertical shear. This is because the smoothing lengths still remain
$ < H$, the disc semi--thickness, beyond that region of the disc which resides
in the equatorial plane of the black hole, and where it
becomes noticeably curved.

\begin{figure}
\epsfig{file=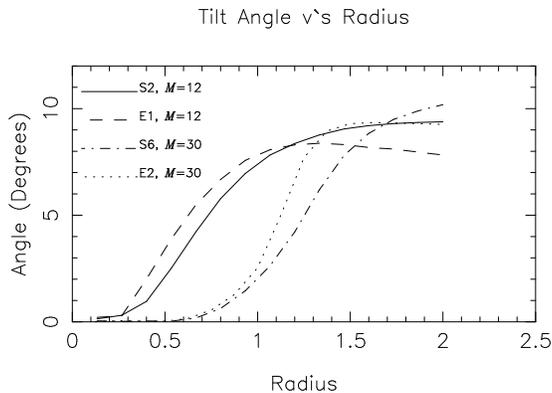, width=\columnwidth}
\caption{This figure shows the effects of including Einstein precession
on the variation of disc tilt with radius. The calculations E1 and E2
have sharper transitions between the inner outer disc relative to their
softened counter parts S2 and S6.}
\end{figure}

We remark that calculation E5, which has the largest disc thickness
of the models we considered with ${\cal M}=5$, showed no signs of alignment
in its inner regions, and represents the limiting disc thickness that allows the
formation of a noticeably warped disc. This has relevance to the properties
of the thick advection dominated accretion discs ({\em i.e.} ADAFs) 
around rotating black holes, which will presumably show no outward signs 
of being warped.

\subsection{Calculations Including the Effects of Einstein Precession} \label{Einstein}

The   discussion of warp   propagation given in  paper I indicates
that efficient, non dispersive bending wave propagation is expected to occur 
only for Keplerian accretion discs in which $\Omega \simeq \kappa$. 
In a situation where $\Omega \ne \kappa$, then the communication of warping
disturbances is expected to become dispersive, with the effects
of this dispersion becoming more important as $| \Omega - \kappa |$ increases
({\em i.e.} as the apsidal precession frequency increases), since the resonance
between the circular and epicyclic motion becomes increasingly detuned.
As was noted in section (\ref{prec-freq}), the frequency of Einstein precession
({\i.e.} advance of periapse) remains relatively high out to quite large
radii in our disc models, indicating that it may be an important 
consideration in determining the structure and evolution of warped
accretion discs around black holes. 

\begin{figure}
\epsfig{file=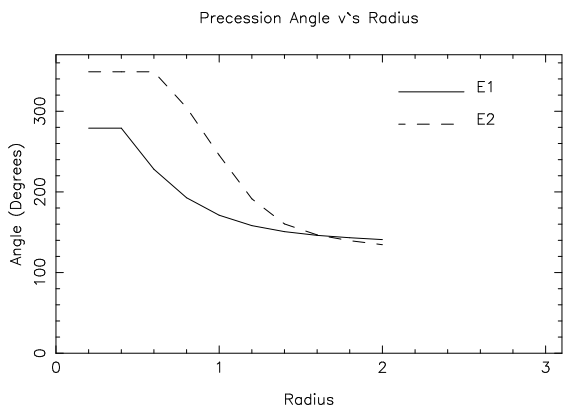, width=\columnwidth}
\caption{This figure shows the degree of twist in the discs at
the end of
calculations E1 and E2. Note that the inner disc is twisted relative to the
outer disc by an angle of $\sim 90$ degrees in calculation E1, and by
$\sim 120$ degrees in E2, which should be contrasted with
the twist in the calculations with gravitational softening S2 and S6.}
\end{figure}

A comparison between the calculations S2 and E1 is presented in Fig. (10). 
The {\em solid} line shows the variation of disc tilt versus radius for calculation
S2, whereas calculation E1 is indicated by the {\em dashed} line. The
primary difference between the cases with Einstein precession included
and those with it neglected is that the transition between the inner
and outer disc planes occurs much more sharply when Einstein precession is
included. The calculations for $i=10^o$ and ${\cal M}=30$ are also shown in
Fig. (10), S6 being shown by the {\em dashed--dotted} line and 
E2 by the {\em dotted} line,
and illustrate this steeper transition between the inner and outer disc 
planes more clearly. This arises because the communication of the disc tilt
between between the two misaligned planes is more effective when 
Einstein precession is neglected, so that the transition may be smoothed
out more effectively by the enhanced diffusion of disc tilt.

One effect of this modification of disc structure by the inclusion of
Einstein precession in our calculations is to change the rate at which
the outer disc tends to align with the equatorial plane of the black hole.
The existence of a more discontinuous transition between the two disc planes 
leads to a larger frictional/viscous interaction between the outer and 
inner disc at
their points of intersection, and thus to a larger torque that acts to realign 
the two separate regions of the disc. Since the inner disc is anchored 
to the black hole equatorial plane by the Bardeen--Petterson effect,
the outer disc is also forced to evolve towards this plane.
Our calculations indicate that this rate of realignment is enhanced 
when Einstein precession is included. Although this seems counter 
intuitive
since disc tilt is supposed to diffuse through the disc
at a lower rate when
the rotation profile is non Keplerian, this result is in agreement with the
predictions of linear theory, as discussed in section (\ref{align}). 

In addition to the transition between the inner and outer disc planes 
being steeper, we also find that the disc twist is greater when Einstein
precession is included. Fig. (11) shows the precession angles of different
disc annuli at the end of the calculations E1 and E2. These plots should be 
compared with Figs. (4) and (9) 
which show the precession angles for the calculations
S2 and S6. We find that calculation E1 results in a twist of the disc
of $\sim 90^o$ between the radii $R=0.6$ to 2.0, which should be compared with 
$\sim 25^o$ for calculation S2. A similar comparison between E2 and S6 yields
twist angles of $\sim 120^o$ and $90^o$, respectively.
This is entirely consistent with linear analysis
[see section (\ref{align})] which indicates that inward
propagating  short
wavelength bending waves should be present in the inner part
of the transition region producing a twist there.
Physically
the large twist angles produced in the simulations
are a result of the diminished efficacy of 
communication between neighbouring annuli in the disc when the rotation profile
becomes significantly non Keplerian. In order for such a disc to settle
to a steady state structure which is in a state of near solid body precession,
the disc must become more severely distorted and twisted so as to increase
the torques acting between neighbouring annuli of gas. 

We remark that  many  oscillations in disc tilt as indicated
by the linear analysis discussed here
and  found in the
linear calculations of Ivanov \& Illarianov (1997), do not appear
in our nonlinear  simulations. This is presumably because non linear effects
lead to the damping of these short wavelength features, and alignment
of the inner disc regions 
in which the 
tilt amplitude
would otherwise change rapidly on small length scales.

\subsection{Effects of Increasing the Inclination Angle $i$}

\begin{figure}
\epsfig{file=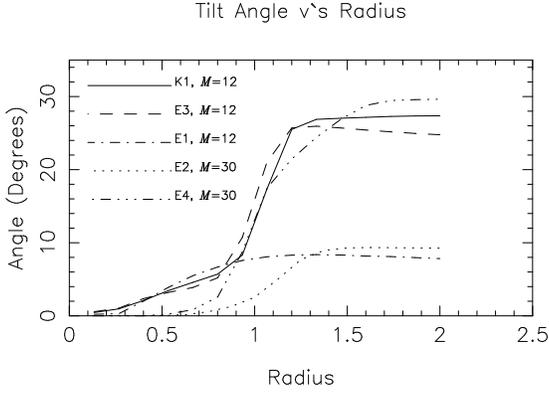, width=\columnwidth}
\caption{This figure shows the variation of disc tilt with radius
for the high
inclination runs E3, E4, and K1. Also plotted are the tilt variation
calculated in the models E1 and E2 for comparison purposes.
Note that the misaligned inner and outer disc parts occupy a similar
radial extent when going from low to high inclination, leading to a sharp
transition in the high inclination cases.}
\end{figure}

Three calculations were performed with larger degrees of inclination
between the black hole equatorial plane and the original disc plane.
These were the calculations listed as E3, E4, and K1 in table~\ref{tab1} .

The results of these calculations are plotted in Fig. (12), along
with those of calculations E1 and E2 which are plotted for the
purpose of comparison. As in the previous runs, the calculations result in a 
disc structure which is essentially composed of two mutually inclined regions,
the inner and outer disc, and a transition zone between them.
In the low inclination ($i=10^o$) runs, the transition zone appears 
to join the 
two misaligned inner and outer disc parts more smoothly than is the
case with the higher inclination ($i=30^o$) runs. This is because the
inner disc, which becomes aligned with the black hole equator, 
and the outer disc, which remains in its original plane, 
occupy approximately the
same intervals in radius irrespective of whether the inclination angle is 
$10^o$ or $30^o$. Thus, the transition between these two planes occurs more
steeply simply because of the greater degree of inclination when $i=30^o$.
The steepness of the transition in the $30^o$ inclination degree runs,
combined with an inspection of the disc structures resulting 
from the calculations
E3 and E4, indicate that the discs are close to breaking into two discrete
pieces with only a tenuous bridge of material connecting them. The
dominant means by which the two separate pieces of the disc communicate
with one another is probably through the viscous coupling at their  region of
 intersection rather than through wave--like
or diffusive communication.

\subsection{A Larger Disc Model}

\begin{figure}
\epsfig{file=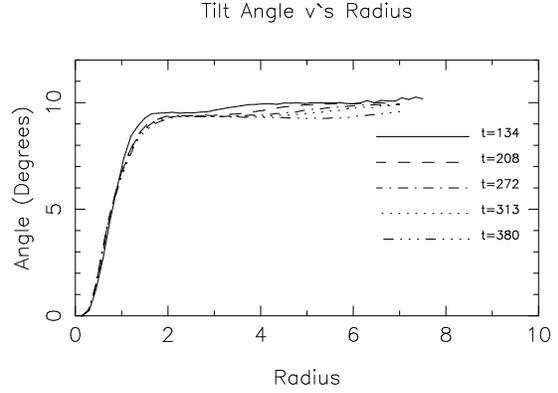, width=\columnwidth}
\caption{\label{fig13} This figure shows the time 
variation of disc tilt with radius
for the large disc model E6 with outer radius at $r=7$ described in the text.
It is apparent that the transition radius is essentially set up
after $t \sim 134$ with some additional relaxation occurring up to
$t \sim 208$. We note that the differential precession 
time at $R=2$ is $\tau_{dp} =166$. After this time the transition radius remains
fixed, with the disc outside of this radius being slowly torqued into alignment
with the equatorial plane of the black hole.}
\end{figure}

A calculation was performed using a disc model with an outer radius at
$r=7$ in order to ensure that the transition radii obtained in 
previous sections are not affected by the smaller disc models used.
This calculation is listed as run E6 in table~\ref{tab1}, and employed
$2 \times 10^5$ particles. The midplane Mach number ${\cal M} \sim 14$, and
the expected value of $\alpha_1 \sim 0.04$, so that bending
waves are expected to travel in this disc model. The effects of Einstein
precession were included so that equation (\ref{Phi2}) was used to calculate 
the gravitational force.

The evolution of the disc tilt for this run can be observed in Fig.
(\ref{fig13}) where each line corresponds to the disc tilt
versus radius at different times during the run. The times are shown
in the figure in units of $\Omega^{-1}$ evaluated at $R=1$. We note that
the warped disc and the transition radius are expected to be set up
after a time corresponding to the differential precession time
at the radius of the transition zone. From Fig. (\ref{fig13}) we
notice that the transition radius, as defined in the first
paragraph of section~\ref{BP-init}, 
is located at $R \sim 0.77 = 20 R_+$, with
the disc tilt approaching that of the original disc beyond a radius of 
$r \sim 2$. This result agrees very well with those obtained for
the disc models with ${\cal M} \sim 12$ and outer radii of $r=2$
described in previous sections, as seen in table~\ref{tab1}. The larger
disc of radius $r=7$ 
is slightly thinner with
${\cal M} \sim 14$, with the result that the transition zone is
slightly further out in this case.
The differential precession time at a radius of
$r=2$ is $\tau_{dp} \simeq 166$, which is the time required for the transition
zone to become fully relaxed.
The calculation ran for a time of $\sim 400$ $\Omega(R=1)^{-1}$.
It is apparent that once the transition zone has been established
on the precession time, its position remains fixed when the calculation
is continued for a number of precession times.  The disc exterior to the
transition zone undergoes some longer time scale readjustment as a
transient disturbance propagates through the disc from the central regions.
As is the case with the calculations presented in previous
sections, the outer disc undergoes a process of slow realignment with
the black hole equator, with the location of the transition radius
remaining fixed throughout this period of the evolution.

\begin{figure*}
\epsfig{file=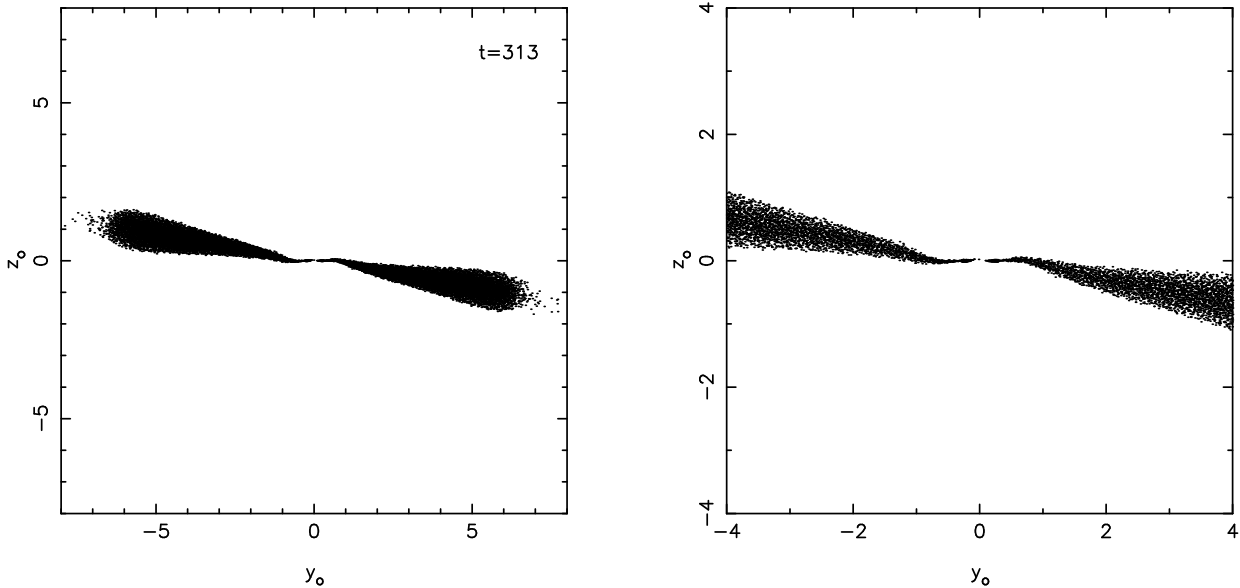, width=\textwidth}
\caption{\label{fig14} This figure shows the structure of the warp
for run E6. A particle projection in the $y$--$z$ plane of a slice 
through the disc is shown in the first panel ilustrating the global
structure of the disc. A close--up view of the warped region is shown in
the second panel. The plots correspond to a time of $t=313 \; \Omega(R=1)^{-1}$.
The differential precession time at the outer edge of the transition zone
($r=2$) is $\tau_p =166 \; \Omega(R=1)^{-1}$ .}
\end{figure*}

Particle projection plots are presented in Fig. (\ref{fig14}) showing the
structure of the warp at a time $t=313 \; \Omega(R=1)^{-1}$, after the warp has
settled to a quasi-steady state. The first panel shows a projection in
the $y$--$z$ plane of a slice through the disc, and presents a global
view of the warp. A closer view of the central regions is shown in the second
panel, and illustrates how the warp and transition zone are confined to the 
central regions of the disc.

\subsection{Alignment Time Scale for Black Hole and Accretion Disc} \label{align}
As the calculations in the previous sections have shown,
the effect of the Lense--Thirring precession induced by the dragging of 
inertial frames by a Kerr black hole is to cause the inner regions of a 
misaligned accretion disc to become aligned with the black hole
angular momentum vector. 
Newton's third law, however, implies that equal but opposite torques are
exerted on the black hole by the disc, causing it's angular momentum
vector, in turn,  to
become aligned with the angular momentum vector of the outer disc material.

This process was considered by Rees (1978), who estimated the alignment
time scale for the black hole and accretion disc by considering the mass
flux through the disc. The direction of the angular momentum of this
infalling matter is changed as it passes through the transition region
between the aligned and misaligned disc, so that a simple estimate of
the alignment time scale may be obtained from

\be \tau_{align} \simeq \frac{J_{BH}}{\dot M j_d(R_{BP}) \sin{(i)}}, \label{talign1}
\ee
where $J_{BH}$ is the black hole angular momentum, $\dot M$ is the mass 
accretion rate through the disc, $j_d(R_{BP})$ is the specific angular
momentum of disc material as it enters the warped region at $R_{BP}$,
 and $i$ is the
inclination angle between the outer disc and black hole equator.
Using this general picture, Rees (1978) calculated the time scale for
alignment in AGN and estimated it to be in the region of $\sim 10^8$ yr, 
comparable
to the ages of jets in AGN whose directions seem to have been constant over 
these time scales.

The above process was considered in more detail by Scheuer \& Feiler (1996).
These authors derived an analytic expression to describe the steady
state warped structure of a disc around a Kerr black hole,
using a linearized set of
equations derived from those of
Pringle (1992), to describe the evolution of a warped disc.
By calculating the Lense--Thirring torque due to the black hole on this disc,
they were able to calculate the reverse torque experienced by the black hole,
and hence to estimate its alignment time scale. They found that 
the more detailed calculations gave time scales in essential agreement 
with the ideas
proposed by Rees (1978), provided that the viscous diffusion coefficients
acting in and out of the disc plane, $\nu_1$ and $\nu_2$, are equal.

These estimates of the alignment time scale were calculated by assuming 
that disc
warp is communicated through a disc on the standard viscous time scale,
so that the transition radius  $R_{BP}$ is then expected to be at 
a large distance from the black hole.  The calculations presented in previous
sections indicate that when the full disc hydrodynamics are considered,
then the evolution of warped discs changes substantially, such that the
transition radius moves closer to the black hole.
This effect has recently been considered by Natarajan \& Pringle (1998),
who have estimated alignment time scales of $\sim 10^5$ yr for black holes
in AGN using 
the formula suggested by Scheuer \& Feiler (1996), but taking into account 
differences between the viscous diffusion coefficients that act
in the plane and perpendicular to the plane of the disc.

The accretion disc models that we calculate may be different from those 
expected
to occur in AGN, since they are relatively thick 
($H/r \sim 0.08$ -- $0.03$), whereas $H/r \sim 10^{-3}$  may be more appropriate to
the outer discs of AGN. In addition, they are of small radial extent. This prevents us from
making detailed calculations of the alignment time scale for black holes in
AGN on the basis of our calculations.
Nonetheless, we can examine the rate at which the global disc tilt changes in
the simulations, and compare it with that expected from the
tilt evolution caused by the advection of misaligned angular momentum
through the warped region, and with the work of
Scheuer \& Feiler (1996). 

If the disc tilt in our simulations changes solely because of advection 
of material with misaligned $J$ through the warped region, where it is
torqued into alignment, then the rate of change of misaligned disc 
angular momentum may be estimated to be
\be \frac{d J}{dt} \simeq 
\dot M j(R_{TS}) \sin{(i)}, \label{J_perp-dot} \ee
where $j(R_{TS})$ is the specific angular momentum of disc material
as it enters the warped region located at $R_{TS}$. 
The corresponding expression that applies when the transition zone occurs
at the Bardeen--Petterson radius, $R_{BP}$ is given by
\be \frac{d J}{dt} \simeq
\dot M j(R_{BP}) \sin{(i)}, \label{J-dot-rees} \ee
where $j(R_{BP})$ is the specific angular momentum of disc material
as it enters the region located at $R_{BP}$.
By calculating $\dot M$ from our simulations, we can estimate the effect of 
these processes, and the associated time scales for them to align the disc
with the black hole and thus by Newton's
third law, the time to align the black hole with the disc. 
We denote these time scales by $\tau_{TS}$ and $\tau_{BP}$,
respectively.

In our finite disc models, we expect the global change in the disc tilt
to arise because of advection of tilt angular momentum through $R_{TS}$,
viscous coupling between misaligned disc annuli, and residual torquing
by the black hole on disc elements that lie somewhat 
beyond $R_{TS}$, which remains
relatively close to the black hole in our simulations.

The time scale for the alignment of a black hole's spin angular momentum
vector with that of the outer part of an accretion disc is given by
Scheuer \& Feiler (1996) to be
\be \tau_{align} \simeq \frac{1}{\pi \Sigma} \left( \frac{a c M}{\nu_2 G}
\right)^{1/2} \label{talign} \ee
where $a$ is the Kerr parameter, $c$ is the speed of light, $M$ is the
black hole mass, $\Sigma$ is the disc surface density, and 
$\nu_2$ is the viscous diffusion coefficient acting on 
the disc warp. We remark that the expression of Scheuer \& Feiler (1996) 
which gives the aligning torque (their equation [9])
indicates that most of the torque contribution arises
from the region of the disc that is 
in the vicinity of the warp or transition radius, 
as was commented upon by these authors themselves, implying that the
alignment time scale obtained is not reliant upon the disc being infinite,
but just significantly larger than the transition radius. 

This is also
the
case  if  the expressions for the tilt (\ref{SFg}) and (\ref{SFg1}),
applicable when $\alpha_1$
is small, 
are 
used   in order to evaluate their expression for the torque.
Then $\nu_2$ in (\ref{talign}) should be replaced such that
\be \nu_2 \rightarrow \nu_2\alpha_1{\left(\sqrt{(\epsilon^2+\alpha_1^2)}
+\epsilon\right)\over\left(\epsilon^2+\alpha_1^2\right)}.\ee
In addition for these cases $\nu_2 = c_B^2/(\alpha_1 \Omega).$  
Note that in the Einstein precession case (\ref{SFg1}),  the limit
$\alpha_1 \rightarrow 0,$ with both $\nu_2$ and $\epsilon$ $\propto 1/r,$
is to be taken  with the result 
that $\nu_2 $ is replaced by a constant value.

An interesting feature of the above is that in the case of
prograde Einstein precession $(\epsilon > 0$) an alignment torque survives
in the limit $\alpha_1 \rightarrow 0,$ whereas in the retrograde
case $(\epsilon <0)$, as occurs with softening, it vanishes.
This is because  of the short wavelength oscillations
or bending waves in the former case. In the latter case
the solution is evanescent and of the wrong phase, having
zero twist,
to produce an alignment torque.
The above suggests that  cases with prograde precession
should show significantly larger twists and  alignment torques
than those with retrograde precession when the viscosity
is small. This is found  to be the case in our simulations. 

The black hole alignment time scale may be written as
\be \tau_{align} \simeq  \left| \frac{dJ}{dt} \right|^{-1}\left({GM^2a \sin(i)\over
c}\right) \label{talign2}
\ee
By Newton's third law, the alignment torque acting on the black hole
is equal and opposite to that acting to align the
disc plane with the equatorial plane of the black hole.
Thus, the alignment torque acting on the disc may be written
\be \frac{dJ}{dt} \simeq  \pi \Sigma \left( \frac{\nu_2 G}{acM} \right)^{1/2}
\left({GM^2a \sin(i)\over
c}\right).
\label{disc-torque} \ee
We note that  $S=G^2M^2 a/c^3$, so that we may write equation
(\ref{disc-torque}) as
\be \frac{dJ}{dt} = \pi \Sigma \left(S \nu_2 G M\right)^{1/2}\sin(i). \label{disc-torque2} \ee
In their discussion, Scheuer \& Feiler (1996)
distinguish between the viscosity coefficient, $\nu_1,$  acting in the plane of the disc
which is responsible for mass flow through the disc, 
and that which 
is responsible for the damping or diffusion of
disc warp, $\nu_2.$

In the situation where the diffusion coefficients are equal
({\em i.e.} $\nu_2 = \nu_1$ where $\nu_1 = \alpha H^2 \Omega$)  
equation (\ref{disc-torque2}) may be written  as
\be \frac{dJ}{dt} \simeq \pi \Sigma (S \nu_1 G M)^{1/2}\sin(i). \label{SF1} \ee

Recent work on the evolution of warped discs, including that presented
in paper I,
indicates that disc warp diffuses on a time scale much faster
than that on which mass diffuses through the disc, and that the
appropriate value for $\nu_2$ is $\nu_2 = \nu_1 / (2 \alpha^2)$.
In order to estimate the alignment time scale for the black hole angular
momentum vector, Natarajan \& Pringle (1998) used this relation
in equation (\ref{talign}). For our purposes of calculating the
torque acting on the disc, using this relation in 
equation (\ref{disc-torque2}) leads to the expression
\be \frac{dJ}{dt} \simeq \frac{\pi \Sigma}{\alpha} \left({S \nu_1 G M\over 2}\right)
^{1/2}\sin(i). 
\label{SF2} \ee

We are able to compare our simulated disc alignment torques
with the values expected from equations (\ref{J_perp-dot}),
(\ref{J-dot-rees}),(\ref{SF1}),
and (\ref{SF2}). We denote the expected time scales for the discs to
align with the black hole equatorial plane calculated from these
equations as $\tau_{TS}$, $\tau_{BP}$,
$\tau_{SF1}$, and $\tau_{SF2}$, respectively.

We denote the time scale for disc alignment
measured from the simulations by $\tau_m.$
We define the time scales for disc alignment to be
\be \tau = | {\bf J}_{\perp} | \left( \frac{dJ}{dt} \right)^{-1} 
\label{general-talign} \ee
where we take ${\bf J}_{\perp}$ to be the perpendicular component
of  the angular momentum of  the disc models at time $t=0.$ 
This value 
is used in the estimates of $\tau_m$, $\tau_{TS}$, 
$\tau_{BP}$, $\tau_{SF1}$, and
$\tau_{SF2}$.

 \begin{table} 
 \begin{center}
 \begin{tabular}{llllllllllllr}  \hline \hline
 Run & $R$ & ${\cal M}$ 
   & $\tau_m$ &
$\tau_{TS}$  & $\tau_{BP}$ & $\tau_{SF2}$ & $\tau_{SF1}$ \\ \hline

&       &    &    &    &   &                       &         \\
S2 & 2  & 12 & 1363 & 11225 & 1348 & 119  & 3826 \\
&       &    &    &    & &                          &         \\
S6 & 2  & 30 & 1628 & 9573 & 1546 & 695 & 4096  \\
\hline 
E1 & 2  & 12 & 378 & 11225 & 1348 &  119 & 3826 \\
&       &     &    &    &         &                 &         \\
E2 & 2  & 30 & 1120 & 10045 & 1570 & 869 & 4392 \\
&       &     &    &    &          &                &         \\
E6 & 7 & 14  & 4000 & 131367 & 18666 & 3219 & 56848 \\
 \hline \hline
 \end{tabular}
 \end{center}
 \caption{\label{tab2}The first column provides the label given to each run, the 
 second column gives the radius. The third column gives the 
 midplane mach number,
 and the fourth, fifth, sixth, seventh, and eighth columns give the measured and
 various analytically predicted alignment time scales, whose symbols are
 defined in the text. Note that calculation labels S$i$ indicate that
 the gravitational potential of the central object was softened, the labels
 E$i$ indicate that no softening was employed but that Einstein precession
 was included.}
 \end{table}

The value of $\tau_m$, the time scale for disc alignment measured from the
simulations, was obtained by measuring the change in  ${\bf J}_{\perp}$
over time, and extrapolating forward to the point when ${\bf J}_{\perp}=0$.

The values of ${\dot M}$ used in equations (\ref{J_perp-dot})
and (\ref{J-dot-rees}), to estimate the values of $\tau_{TS}$ and
$\tau_{BP}$, were obtained from the simulations by measuring the
mass flux through the disc models. The values of $R_{TS}$ were taken from
table~\ref{tab1} and the values of $R_{BP}$ were calculated using equation
equation (\ref{r_bp}). The values of $\nu_1$ and $\alpha$ used in equations 
(\ref{SF1}) and (\ref{SF2}) were obtained from the measurements of
${\dot M}$ and assuming a steady state such that ${\dot M}= 3 \pi \nu \Sigma$.
The values of $\alpha$ used in the calculation of the alignment
time scales for each model were (S2 -- $\alpha=0.022$; S6 -- $\alpha=0.12$;
E1 -- $\alpha=0.022$; E2 -- $\alpha=0.14$; E6 -- $\alpha=0.04$). These are very
similar to the values of $\alpha_1$ obtained from the bending wave calculations
presented in paper I, and indicate that SPH produces a reasonably isotropic
viscosity when applied to this problem.

The alignment time scales calculated from the simulations and the analytical
torque estimates are presented in table~\ref{tab2}.
First of all, we consider the runs that employed softening of the gravitational
potential, since in this case it is relevant to compare these calculations
with the analytical estimates of the alignment time scales, given 
that the softening plays only a minor role in the simulations.
The physical models that lead to the 
analytical estimates of the alignment time scales do not include
the effects of Einstein precession, and so strictly speaking they should not
be compared with the results of calculations E1, E2, and E6 in which Einstein
precession is included and plays an important role.

From table~\ref{tab2} we see that the results for the runs S2 and S6 indicate 
that the predicted
alignment time scale, $\tau_{TS}$, estimated by assuming that
material is only torqued into alignment when it accretes through the transition
zone, $R_{TS}$, is too long, indicating that the viscous torques acting between the
aligned and misaligned disc components, and the residual torque due to the 
black hole beyond $R_{TS}$, are having a substantial effect on the evolution of
the disc tilt. The alignment time scale $\tau_{SF2}$  leads to a prediction
that the discs will align with the black hole equatorial plane on too short a 
time scale, whereas the estimate $\tau_{SF1}$  predicts too long
a time scale. Interestingly, it appears that the estimate $\tau_{BP}$
based on equation (\ref{J-dot-rees}) provides the best estimate of the alignment
time scale in the cases S2 and S6.

The physical picture suggested by Rees (1978) is
of a warped
disc with constant mass
flow in which the misaligned component of the disc material's
angular momentum is transferred to the black hole as it passes through the
warped region located at $R_{BP}$.

It  may seem surprising  that we obtain results
for the alignment time scale that are consistent with Rees (1978)
in view of the facts that
the transition radius is much smaller in our case
and (\ref{SF2}) predicts a significantly faster rate.
But we note that (\ref{SF2}) is based on a linearized approximation
which does not allow for the torques applied in the warped region
of the disc to feedback into the disc structure. One would expect
that this feedback would clear
material  with misaligned angular momentum from
this region which would then have
to be resupplied by mass accretion. This would
result in a dependence of the alignment rate on the mass accretion rate.

From the discussion presented in section~\ref{BP-rad}, we would expect that 
the alignment time scales obtained when Einstein precession is included would be
noticeably shorter than when it is neglected. This arises because of the
larger twist in these cases, leading to a larger torque. By examining the
values of $\tau_m$ listed in table~\ref{tab2} we see that this is indeed the
case. Comparing the runs S2 and E1, we see that the alignment time scale is 
shorter by a factor of $\sim 4$ when Einstein precession is included.
Comparing runs S6 and E2, we observe the same trend, but at a reduced level
because the transition zone is further from the black hole in these cases
and the effects of Einstein precession are reduced accordingly.

By comparing the results of runs E1 and E6, we notice that $\tau_m$ is larger 
than in the latter case because of the larger disc size.
Although the disc model parameters for these two cases differ slightly, 
giving slightly
different transition radii, we would expect that the ratio of
alignment time scales would be similar to the ratio of the initial disc
angular momenta if the aligning torques are similar.
We find that $\tau_m(E1)/\tau_m(E6) \sim 0.1$ and
$J_{\perp}(E1)/J_{\perp}(E6) \sim 0.12$, suggesting that we not
only calculate the position of the transition zone correctly with
our smaller disc models with outer radii $r=2$, but that we also
accurately calculate the alignment torque between the outer and inner discs
since this arises from the transition zone.

\section{Discussion and Conclusion}
In this paper we have performed simulations of accretion discs 
in orbit around a
rotating black hole, which had its spin vector initially misaligned
with that of the disc, in the lowest order post
Newtonian approximation. Discs with midplane Mach numbers between 5 and 30
were considered. For discs with Mach numbers of 5, 9, and 12, the effective
viscosity acting through the $r$--$z$ component of the viscous stress, 
$\alpha_1$, is
such that the warps are controlled by bending waves (see paper I).
For the higher Mach number
of 30, the warps evolve diffusively, but with an associated diffusion 
coefficient that is a factor of $\sim 1/(2 \alpha_1^2)$ larger than that
associated with mass flow through the disc (Papaloizou \& Pringle 1983). 

As expected the central portions 
of the disc models became aligned with the equatorial plane of the black hole
out to a transition radius, beyond which the discs remained close to their
original
plane. This structure was established on the sound crossing,
or warp diffusion, time. This period of initial relaxation to a 
well defined warped disc structure was followed by a period of 
evolution occurring on a longer time scale, corresponding to 
solid body precession of the disc and a slow process of alignment of the outer
disc with equatorial plane of the black hole. Both of these are caused by the
fact that we are forced to consider isolated discs of finite radii for reasons
of numerical tractability. Nonetheless the position of the transition
between the aligned and misaligned disc components is not affected by this.
Because of the much more effective communication of warps than
implied by the standard viscous time scale, the transition radius
was found to be much smaller than that given by Bardeen \& Petterson (1975),
ranging between 15 and 30 gravitational radii  for a hole 
with Kerr parameter $a=1.$

In the warped regions of the disc with changing inclination, the expected
vertical shear (Papaloizou \& Pringle 1983) was seen in the simulations.
We found that these velocities were limited to be subsonic, presumably
by nonlinear effects such as shocks.
An interesting issue is the the stability of this shear flow. 
Parametric instability
associated with bending waves is a possibility (Papaloizou \& Terquem 1995;
Gammie, Goodman, \& Ogilvie 1998, private communication).
This may produce a level of turbulence in a disc with low viscosity.
But it should be noted that the SPH simulations presented here are 
for discs that are already viscous, such that the growth rate of any
instability is exceeded by the viscous damping.
Simulations  with a viscosity small enough to investigate
shear flow instability require too many particles to be practicable at present.

Models were considered both at high and low inclination. In the nonlinear 
high inclination case, although the transition radius
remained approximately in the same location, the transition was more abrupt
than in the low inclination case, indicative of
a tendency for the outer part of the disc to become disconnected
from the inner part. The implications of this are that a disc which
is forced to maintain a nonlinear warp due to severe misalignment
will tend to break into two or more disconnected pieces, rather than
maintain a smoothly warped structure in which the perturbed horizontal
motions remain transonic.

Although the calculations
performed here are in a very different physical regime to that expected for
discs around AGN, the time scale we obtain
for black hole -- disc alignment
is in essential agreement with the ideas of Rees (1978) 
and accordingly the later work of  
Scheuer \& Feiler (1996),  if the viscous diffusion coefficients
acting in and out of the disc plane are taken to be equal.   
We also found that this alignment time scale could be significantly shortened
by the inclusion of Einstein precession, in line with the predictions of a
linear analysis.

\newpage

\end{document}